\documentclass[12pt]{article}

\usepackage{graphicx, color, url}
\usepackage{authblk}

\usepackage{amsmath}
\usepackage{dsfont}
\usepackage{amssymb}
\usepackage{epstopdf}
\usepackage{boxedminipage}
\usepackage{amsfonts}
\usepackage{url}
\usepackage{amsbsy}
\usepackage{setspace}
\usepackage{natbib}
\usepackage{bm}
\usepackage{wasysym}
\usepackage{multirow}
\usepackage{textcomp}
\usepackage{amsthm}
\usepackage{verbatim}
\usepackage{amssymb}
\usepackage{amsmath}
\usepackage{amsbsy}
\usepackage{setspace}
\usepackage{natbib}
\usepackage{bm}
\usepackage{wasysym}
\usepackage{textcomp}
\usepackage{amsmath,amssymb,amsfonts, amsbsy, epsfig, subfig,natbib,soul,hyperref, mathtools}
\usepackage{mathrsfs}
\usepackage{algpseudocode}
\usepackage{algorithm}
\usepackage{graphicx, color, url}
\usepackage{amsmath}
\usepackage{enumitem}

\renewcommand{\theequation}{\thesection.\arabic{equation}}

\newtheorem{theorem}{Theorem}
\newtheorem{corollary}{Corollary}
\newtheorem{proposition}{Proposition} 
\newtheorem{lemma}{Lemma}
\newtheorem{example}{Example} 
\newtheorem{remark}{Remark}
\newtheorem{definition}{Definition}

\newcommand\X{\bm{X}}

\newcommand\sgn{\operatorname{sgn}}

\newcounter{rcnt}[section]

\newcommand{\ntt}{\frac{n_1n_2}{n_1+n_2}}
\newcommand{\nt}{\frac{2n_1n_2}{n_1+n_2}}
\newcommand{\nn}{\frac{n_1+n_2}{n_1n_2}}
\newcommand{\ninv}{\frac{(n_1+n_2)}{2n_1n_2}}
\newcommand{\hQ}{\widehat{Q}}

\newcommand{\Tor}{T}

\usepackage[svgnames]{xcolor}
\usepackage{listings}

\newcommand{\beq}{\begin{equation}}
	\newcommand{\eeq}{\end{equation}}
\newcommand{\beas}{\begin{eqnarray*}}
	\newcommand{\eeas}{\end{eqnarray*}}
\newcommand{\bea}{\begin{eqnarray}}
	\newcommand{\eea}{\end{eqnarray}}
\newcommand{\bei}{\begin{itemize}}
	\newcommand{\eei}{\end{itemize}}
\newcommand{\ben}{\begin{enumerate}}
	\newcommand{\een}{\end{enumerate}}
\newcommand{\bet}{\begin{theorem}}
	\newcommand{\eet}{\end{theorem}}
\newcommand{\bel}{\begin{lemma}}
	\newcommand{\eel}{\end{lemma}}
\newcommand{\bep}{\begin{proposition}}
	\newcommand{\eep}{\end{proposition}}
\newcommand{\bed}{\begin{definition}}
	\newcommand{\eed}{\end{definition}}
\newcommand{\bec}{\begin{corollary}}
	\newcommand{\eec}{\end{corollary}}
\newcommand{\bex}{\begin{example}}
	\newcommand{\eex}{\end{example}}

\def\0{\boldsymbol{0}}

\def\X{\boldsymbol{X}}

\def\Y{\boldsymbol{Y}}

\def\W{\boldsymbol{W}}

\newbox\TempBox \newbox\TempBoxA

\def\Var{\textsf{Var}} 

\newcommand{\EE}{\mathbb{E}}
\newcommand{\II}{\mathbb{I}}
\newcommand{\PP}{\mathbb{P}}

\def\0{\boldsymbol{0}}

\def\X{\boldsymbol{X}}
\def\Y{\boldsymbol{Y}}

\def\W{\boldsymbol{W}}

\definecolor{gray}{RGB}{150,150,150}
\definecolor{blue}{RGB}{000,000,200}
\definecolor{green}{RGB}{000,150,100}
\definecolor{purple}{RGB}{200,000,180}

\begin{document}
	\title{A Locally Adaptive Shrinkage Approach to False Selection Rate Control in High-Dimensional Classification} 
	\author{Bowen Gang$^1$, Yuantao Shi$^2$, and Wenguang Sun$^3$}
	\date{}
	
	\maketitle


	\maketitle
	\begin{abstract}
		%
		The uncertainty quantification and error control of classifiers are crucial in many high-consequence decision-making scenarios. We propose a selective classification framework that provides an ``indecision'' option for any observations that cannot be classified with confidence. The false selection rate (FSR), defined as the expected fraction of erroneous classifications among all definitive classifications, provides a useful error rate notion that trades off a fraction of indecisions for fewer classification errors. We develop a new class of locally adaptive shrinkage and selection (LASS) rules for FSR control in the context of high-dimensional linear discriminant analysis (LDA). LASS is easy-to-analyze and has robust performance across sparse and dense regimes. Theoretical guarantees on FSR control are established without strong assumptions on sparsity as required by existing theories in high-dimensional LDA. The empirical performances of LASS are investigated using both simulated and real data.  
		
	\end{abstract}
	
	\vspace{9pt}
	\noindent {\it Key words and phrases:}
	Classification with confidence, False discovery rate,  Linear discriminant analysis, Risk control, Shrinkage estimation.
	\footnotetext[1]{Department of Statistics and Data Science, Fudan University.}  
	\footnotetext[2]{Department of Statistics, University of Chicago.}  
	\footnotetext[3]{Center for Data Science, Zhejiang University. } 
	
	\thispagestyle{empty}

	\newpage
	\renewcommand{\theequation}{\thesection.\arabic{equation}}

	\setcounter{section}{0} 
	\setcounter{equation}{0} 

	\setcounter{page}{1} 
	
	\section{Introduction}
	
	Linear discriminant analysis (LDA) has been widely used in classification problems. We focus on the basic setup, which assumes that the observations are $p$-dimensional vector-valued features that are drawn with equal probability from one of the two multivariate normal distributions:
	\beq\label{model}
	\mbox{$\mathcal N(\pmb\mu_1, \Sigma)$ (class 1) and $\mathcal N(\pmb\mu_2, \Sigma)$ (class 2)}.
	\eeq
	Let $\pmb W\in \mathcal R^p$ be a new observation. Denote $\pmb\mu=\frac{\pmb\mu_1+\pmb\mu_2}2$ and $\pmb d=\pmb\mu_1-\pmb\mu_2$. The procedure that achieves the minimal misclassification risk is Fisher's linear discriminant rule:
	\beq\label{LDA-rule}
	\delta^F=\mathbb I\left\{(\pmb W-\pmb\mu)^\top\Sigma^{-1}\pmb d<0\right\} + 2\cdot \mathbb I\left\{(\pmb W-\pmb\mu)^\top\Sigma^{-1}\pmb d\geq 0\right\},
	\eeq 
	which assigns $\pmb W$ to class $c$ if $\delta^F=c$, $c=1, 2$. When $\pmb\mu_1$, $\pmb\mu_2$ and $\Sigma$ are unknown, the common practice is to 
	construct a data-driven LDA rule by obtaining suitable estimates of the unknown quantities in \eqref{LDA-rule}. In the high-dimensional setting, naive sample estimates become highly unstable, and a plethora of regularized LDA rules have been proposed 
	and shown to achieve substantial improvements in prediction accuracy (\citealp{Fri89, Tibetal03, WitTib09, cai2011direct, shao2011sparse, Maietal12, tony2019high}; among others). However, it still remains unknown how to assess the uncertainties and control the decision errors in high-dimensional LDA. This article proposes a selective classification approach to controlling the false selection rate (FSR). We develop a new class of data-driven LDA rules based on locally adaptive shrinkage and selection (LASS), and illustrate how LASS can be deployed in decision-making scenarios to control the FSR at a user-specified level.

	\subsection{Selective classification and false selection rate}
	
	Uncertainty quantification and error control are crucial in many sensitive decision-making scenarios. The decision errors, which can be very expensive to correct, are often unavoidable due to the intrinsic ambiguity of a classification problem. Consider the ideal setting where the multivariate normal parameters $\pmb\mu_1$, $\pmb\mu_2$ and $\Sigma$ are known. Then among all classification rules, the LDA rule \eqref{LDA-rule} achieves the minimum classification risk $1-\Phi\left( \dfrac{1}{2} \sqrt{\pmb d^\top \Sigma^{-1}\pmb d}\right)$, where $\Phi(\cdot)$ is the cumulative distribution function (CDF) of a standard normal variable. However, this minimum  risk can still be unacceptably high when the signal to noise ratio $\sqrt{\pmb d^\top \Sigma^{-1}\pmb d}$ is low. The issue is exacerbated in practice, particularly in high-dimensional settings, where we must employ ``plug-in'' rules learned from limited training data. 
	
	In contrast with conventional classification algorithms which are forced to make classifications on all new observations, a useful strategy for uncertainty quantification involves providing an \textit{indecision} option (also referred to as abstention or reject option) for any observations which cannot be classified with confidence. The observations with indecisions can then be separately evaluated. This strategy is attractive in practice when the cost of handling indecisions is less than that of fixing a classification error. To see how it aligns with the social and policy objectives, consider a high-consequence classification scenario where one needs to assess the likelihood of a defendant becoming a recidivist. Obviously the social cost of incorrectly classifying a low-risk individual as a recidivist is much higher than that of an indecision -- – it is worthy of waiting and collecting additional contextual knowledge of the convicted individual to mitigate the ambiguity. Likewise, in medical screening, a misclassification can result in either missed medical care or unnecessary treatments, both of which can be much more expensive than turning the patient over for a more careful examination/evaluation. 
	
	Suppose we observe labeled training data $\mathcal D^{train}$.
	The goal is to predict the classes for $m$ new observations $\mathcal D^{test}=\{\pmb W_j: 1\leq j\leq m\}$. This article considers a \emph{selective classification} framework that only makes definitive decisions on a \emph{selected subset} of $\mathcal D^{test}$, while the remaining subjects will receive indecisions (i.e. be rejected for further investigation). The reject/indecision option, which is much less expensive to handle, is considered as a wasted opportunity rather than a severe error. We propose to control the false selection rate (FSR), which is the expected fraction of erroneous classifications among all definitive classifications. Selective classification with FSR control provides an effective approach to uncertainty quantification and error control. We demonstrate that with the reject/indecision option, the FSR can be controlled at a user-specified level. When the signal to noise ratio is low, the degree of ambiguity in the classification task can be, in a sense, captured by the fraction of indecisions in $\mathcal D^{test}$. Hence, a more powerful data-driven rule, subject to the FSR constraint, translates to a smaller fraction of indecisions, which means that less wasted efforts are needed to perform separate evaluations. 
	
	\subsection{FSR control via locally adaptive shrinkage and selection}
	
The task of controlling the FSR in high-dimensional LDA is challenging; we start by discussing several limitations of existing works. 
	
	First, the methodology and theory of many high-dimensional LDA rules (e.g. \citealp{cai2011direct, shao2011sparse, Maietal12, tony2019high}) critically depend on strong sparsity assumptions, which may not hold in practice. The sparsity assumption is counter-intuitive from the perspective of classification error control. Consider the simple case where all non-zero coordinates in $\pmb d=\pmb\mu_1-\pmb\mu_2$ take the same value. Then a larger $l_0$ norm of $\pmb d$ (i.e. non-sparse setting) virtually implies that the two classes are better separated, and hence, the control of classification risk should become easier. However, many state-of-the-art LDA rules lack theoretical justifications and often do poorly in the supposedly easier non-sparse setting (Section \ref{simu}). Second, the analysis of the error rate of a classifier often requires a precise quantification of the quality of its outputs, which is in general intractable due to the complexity of contemporary LDA rules. Finally, most learning algorithms are driven by the need of improving prediction performance instead of avoiding high-consequence decision errors. It is unclear how to tailor existing algorithms to trade off a fraction of indecisions for fewer classification errors, and further, how to calibrate suitable data-driven thresholds to control the FSR at a user-specified level. 
	
	This article develops a class of FSR rules via a locally adaptive shrinkage and selection (LASS) algorithm. LASS consists of three steps: first estimating a score according to the LDA rule \eqref{LDA-rule}, second ordering all individuals according to the estimated scores, and finally choosing upper and lower data-adaptive thresholds to select individuals into the two classes, with the unselected ones assigned to the \emph{indecision} group. We prove theories to establish the asymptotic validity of LASS for FSR control under mild conditions. A key innovation in our method is the construction of intuitive and easy-to-analyze shrinkage factors that are capable of reducing uncertainties without strong assumptions on sparsity. LASS provides a principled and theoretically solid LDA rule that has comparable performance with state-of-the-art classification rules (e.g. \citealp{cai2011direct, shao2011sparse, tony2019high}) in the sparse setting and substantially better performance under the non-sparse setting. The theoretical adaptiveness of LASS to the unknown sparsity and its robust numerical performance across sparse and dense settings are attractive for practitioners -- particularly in real world applications we only ``bet on sparsity''; this working assumption (of sparsity) can distort the hardness of the problem, and hence, lead to wrong choices of method.

	\subsection{Our contributions}
	
	Our work makes several contributions. First, selective classification via FSR control provides a useful approach to risk-sensitive  decision-making scenarios, where classification errors have high impacts on one's social, economic or health status. Second, we develop a novel shrinkage rule for estimating the linear discriminant score, which is effective for reducing the uncertainties in high dimensions. The estimator is intuitive, assumption-lean, easy-to-analyze and enjoys strong theoretical properties. Third, we derive data-adaptive decision boundaries based on the shrunken LDA rule to select and classify the observations. Theoretical guarantees on FSR control are established under much weaker conditions compared to existing theories on sparse LDA. 
	
	\subsection{Related works}
	
	We discuss several related lines of research to further explain the merits of LASS  and place our contributions in context. 
	
	Under the high-dimensional sparse setting, \cite{bickel2004some} demonstrated that the naive Fisher's rule can perform no better than random guess. A plethora of regularized LDA rules have been proposed to exploit the sparse structure in the data; a few representative works include the shrunken centroid method \citep{Tibetal03}, the LPD and AdaLDA rules \citep{cai2011direct, tony2019high}, and other penalized or thresholding methods \citep{shao2011sparse, Maietal12}. However, as we demonstrate in our numerical studies, these methods do not work well under the non-sparse setting. LASS, which employs an adaptive shrinkage rule with robust performance \emph{across sparse and dense regimes}, is provably valid for error rate control. 
	
	Exemplified by the James-Stein estimator \citep{james1992estimation} and Tweedie's formula (e.g. \citealp{BroGre09, Efr11, KoeMiz14}), shrinkage is a powerful and ubiquitous idea in compound estimation. Under the independence assumption (i.e. $\Sigma$ is a diagonal matrix), the implementation of the LDA rule \eqref{LDA-rule} boils down to the compound estimation of $\pmb \mu$ and $\pmb d$. \cite{efron2009empirical}, \cite{greenshtein2009application} and \cite{dicker2016high} proposed empirical Bayes (EB) methods (Tweedie's formula and g-estimation) to construct ``plug-in'' LDA rules. EB shrinkage can effectively reduce the uncertainty in high dimensions without the sparsity assumption. However, there are several drawbacks. First, existing EB rules ignore the correlations, which may lead to suboptimal shrinkage factors and hence inferior LDA rules. By contrast, LASS performs shrinkage in a coordinate-wise shrinkage manner, which enjoys strong numerical and theoretical properties under dependence. Second, in contrast with EB ``plug-in'' rules that are rather complicated to analyze, the uncertainty quantification of LASS is simple, which enables the development of data-driven rules and theory on FSR control. 
	
	\subsection{Organization and notations}
	
	The article is organized as follows. Section \ref{Prob-form:sec} presents the problem formulation and derives the oracle rule for FSR control. The data-driven LASS is developed in Section \ref{estimation:sec} with its theoretical properties established in Section \ref{theory:sec}. The results are presented in Section \ref{simu}. Proofs are relegated to the Supplementary Material. 
	
	Summary of notation. Denote $\pmb{\mu}=\frac{\pmb{\mu}_1+\pmb{\mu}_2}2$, $\pmb{d}=(d_1,...,d_p)^\top=\pmb{\mu}_1-\pmb{\mu}_2$, $\bar{\pmb{X}}=(\bar{X}_1,\ldots,\bar{X}_p)^\top=(1/n_1)\sum_{i=1}^{n_1}\pmb{X}_i$ and  $\bar{\pmb{Y}}=(\bar{Y}_1,\ldots,\bar{Y}_p)^\top=(1/n_2)\sum_{i=1}^{n_2}\pmb{Y}_i$. $I_p$ denotes the $p\times p$ identity matrix. For matrix $A$ and $1\le w\le \infty$, the matrix $l_w$ norm is defined as $||A||_w=$sup$_{|\pmb x|_w\le1}|A\pmb x|_w$. The largest/smallest eigenvalue of $A$ is denoted $\lambda_{max}(A)$/$\lambda_{min}(A)$. 

	\setcounter{equation}{0} 
	
	
	\section{Problem Formulation}\label{Prob-form:sec}
	
	This section first introduces a generalized discriminant rule, then defines the false selection rate, and finally outlines the roadmap for methodological developments.
	
	\subsection{The generalized discriminant rule}
	
	Let $\pmb W$ be a new observation and $S\coloneqq S(\pmb W)$ be a generic score, with larger (smaller) $S$ indicating a higher chance of being in class 2 (class 1). 
	Suppose we need to classify $m$ new observations $\{\pmb W_j: 1\leq j\leq m\}$, which are drawn with equal probability from \eqref{model}. It is natural to consider the following generalized discriminant rule $\pmb\delta=(\delta_j: 1\leq j\leq m)$, where
	\beq\label{LDA-rule2}
	\delta_j=\mathbb I\left\{S(\pmb W_j)<t_l\right\} + 2\cdot \mathbb I\left\{S(\pmb W_j)\geq t_u\right\}, \quad 1\leq j\leq m.
	\eeq 
	In the above, $t_l$ and $t_u$ represent the lower and upper thresholds, respectively, with the requirement that $t_l\leq t_u$. A key difference between the two discriminant rules \eqref{LDA-rule2} and \eqref{LDA-rule} is that \eqref{LDA-rule} uses $t_l=t_u=0$ whereas the generalized rule \eqref{LDA-rule2} allows $t_l<t_u$. The interval $(t_l, t_u)$ is called an \emph{ambiguity region}. The true class of any observation falling within this region cannot be determined with confidence. The values of $t_l$ and $t_u$ will be determined according to user-specified error rates, which will be discussed in the next subsection. It follows that $\delta_j$ defined in \eqref{LDA-rule2} can take three possible values in the action space $\mathcal A=\{0, 1, 2\}$, with $\delta_j=k$ indicating that we classify $\pmb W_j$ to class $k$, $k=1, 2$, while $\delta_j=0$ indicating that we make an indecision or rejection option (\citealp{herbei2006classification, SunWei11, lei2014classification}). Denote  $\{\theta_j: 1\leq j\leq m\}\in\{1, 2\}^m$ the unknown true classes. Consider the example of medical screening, where $\theta_j=1$ ($\theta_j=2$) indicates that the patient is healthy (sick). Then a patient with $\delta_j=1$ will not receive the treatment, with $\delta_j=2$ will receive the treatment, with $\delta_j=0$ will be followed up for further evaluation.
	
	\subsection{False selection rate}
	
	In risk-sensitive applications, we view misclassifications as severe errors, the fraction of which should be controlled at a low level. Under the selective inference framework (\citealp{Ben10}), the error rate is defined to assess the quality of the selected subset, in which observations receive definitive classifications. By contrast, the indecisions are viewed as wasted opportunities and used to describe the power notion. 
	
	For the binary setting, we may encounter two types of misclassifications: $(\theta=1, \delta=2)$ and $(\theta=2, \delta=1)$. If the two directions are symmetric, it is natural to consider the false selection rate (FSR) without distinguishing the two types: 
	\beq\label{FSR}
	\mbox{FSR}=\EE\left[\frac{\sum_{j=1}^m \II\left(\theta_j\neq \delta_j, \delta_j\neq 0\right)}{\left\{\sum_{j=1}^m \II(\delta_j\neq 0)\right\} \vee 1 } \right],
	\eeq
	where $x\vee y=\max(x,y)$. The FSR was recently considered in \cite{Ravetal21} in a different context (fairness in machine learning), and is analogous to the false discovery rate (\citealp{BenHoc95}) in one-class classification problems (outlier detection) \citep{Batetal21, Angetal21}. The FSR reduces to the misclassification rate $\frac 1 m \EE\{\sum_{j=1}^m (\theta_j\neq\delta_j)\}$ if indecisions are not allowed (i.e. $\delta_j\neq 0$ for every $1\leq j\leq m$).  
	
	In the asymmetric situation we may define the class-specific FSR:
	\beq\label{FSR-k}
	\text{FSR}^c= \mathbb{E}\left[   \dfrac{ \sum_{j=1}^{m}\mathbb{I}(\delta_j=c,\theta_j\neq c ) }{ \{ \sum_{j=1}^{m}\mathbb{I}(\delta_j=c)     \}\vee 1    }    \right], \quad c=1, 2. 
	\eeq
	This provides a useful error notion in applications where one type of error is more sensitive than the other and it is desirable to set different tolerance levels for the two types of errors\footnote{The class-specific $\mbox{FSR}^c$ is connected to but fundamentally different from the Neyman-Pearson classification framework \citep{ScoNow05, RigTon11} for asymmetric error control. The class-specific $\mbox{FSR}^c$, as a concept under the selective inference framework, is analogous to the FDR in multiple testing, whereas Neyman-Pearson classification operates under the classical Type I/II error paradigm in single hypothesis testing. Moreover, the two lines of research focus on substantially different issues.}. To this end, we focus on the setup allowing class-specific constraints: $\mbox{FSR}^c\leq \alpha_c$, $c=1, 2$. As a special case, one may set $\alpha_1=\alpha_2=\alpha$. If the class-specific constraints $\mbox{FSR}^c\leq \alpha$ are fulfilled for both $c=1$ and $c=2$, then it can be shown that the global constraint $\mbox{FSR}\leq \alpha$ defined in \eqref{FSR} is also fulfilled. 
	
	The selective classification framework enables one to control the FSR at a user-specified level, which can be impossible without the indecision option. However, the price we need to pay is the wasted opportunity for performing separate evaluations on the indecisions. The user-specified error bounds {$\alpha_c$} reflect our tolerance levels of the associated risks. To simultaneously quantify the degree to which the decisions can be trusted and minimize the wasted efforts, we consider a constrained optimization problem. Let $\text{ECC}=\mathbb{E}\left\{\textstyle\sum_{j=1}^{m}\mathbb{I}(\theta_j=\delta_j) \right\}$ denote the expected number of correct classifications. The goal is to
	\beq\label{ECC}
	\mbox{maximize the ECC subject to $\mbox{FSR}^c\leq \alpha_c$, $c=1,2$. } 
	\eeq
	
	\subsection{Oracle rules for FSR control}\label{sec3}
	
	
	In this subsection we derive a class of oracle FSR rules. To motivate our methodology, consider an asymptotically equivalent error rate (Appendix \ref{fsrmsfr}), the marginal FSR
	\beq\label{mFSR-k}
	\text{mFSR}^c= \dfrac{ \mathbb{E}\left\{\sum_{j=1}^{m}\mathbb{I}(\delta_j=c,\theta_j\neq c )\right\} }{\mathbb{E} \left\{\sum_{j=1}^{m}\mathbb{I}(\delta_j=c) \right\}}. 
	\eeq  
	
	We aim to develop a selective classification rule that solves the following constrained optimization problem: maximize the ECC subject to $\mbox{mFSR}^c\leq\alpha_c$, $c$=1, 2.	Next we prove an intuitive result that the optimal mFSR rule is a thresholding rule based on the optimal LDA function $S_j^\pi\equiv(\pmb W_j-\pmb\mu)^\top\Sigma^{-1}\pmb d$ (or its monotone transformations). 
	
	Consider a generalized discriminant rule $\pmb\delta(t_1, t_2)=(\delta_j: 1\leq j\leq m)$ of the form \eqref{LDA-rule2}:
	$\delta_j=\mathbb{I}\left(1-{T}^j<t_1\right)+2\mathbb I\left({T}^j<t_2\right)$,  $1\leq j\leq m,$
	where 
	$${T}^j\coloneqq T(\W_j)=\PP(\theta_j=1|\W_j)=\dfrac{\exp(S_j^\pi)}{\exp(S_j^\pi)+1},$$ and $t_1, t_2\in(0,1)$ are the lower and upper thresholds satisfying $t_1<1-t_2$. As ${T}^j$ is a monotone transformation of $S_j^\pi$, generalized LDA rules based on ${T}^j$ and $S_j^\pi$, with suitably adjusted thresholds, are equivalent. We use ${T}^j$ instead of $S_j^\pi$ to facilitate the development of a step-wise algorithm, which is described at the end of this section.

	Let $Q^c(t_c)$ be the $\text{mFSR}^c$ of $\pmb\delta(t_1, t_2)$, $c=1, 2$. Define the oracle thresholds
	$
	t^c_{OR}=\sup\left\{t: Q^c_{}(t)\leq \alpha_c\right\}, c=1, 2. 
	$
	To avoid assigning an individual to multiple classes, we assume that $\alpha_1$ and $\alpha_2$ have been chosen such that both $t^1_{OR}$ and $t^2_{OR}$ are less than {or equal to} 0.5\footnote{This is an assumption to facilitate theoretical development. If there is overlapping selection in practice, we can simply classify the individual to the class with larger class probability $P(\theta_j=c|\W_j), c=1, 2.$}. 
	Define the oracle mFSR procedure $\pmb \delta_{OR}=(\delta_{OR}^j: 1\leq j\leq m)$, where
	\begin{equation}\label{oracle-rule}
		\delta_{OR}^j=\mathbb{I}\left(1-{T}^j<t^1_{OR}\right)+2\cdot\mathbb I\left({T}^j<t^2_{OR}\right).
	\end{equation}
	The next theorem shows that $\pmb\delta_{OR}$ is optimal.
	
	\begin{theorem} \label{thm:optimal_or} 
		Let $\mathcal D_{\alpha_1,\alpha_2}$ be the collection of all classification rules such that for any $\pmb\delta\in\mathcal D_{\alpha_1,\alpha_2}$, $mFSR^1_{\pmb\delta}\leq \alpha_1$ and $mFSR^2_{\pmb\delta}\leq \alpha_2$. Then $\mbox{ECC}_{\pmb\delta}\leq \mbox{ECC}_{\pmb\delta_{OR}}$ for any $\pmb\delta\in \mathcal D_{\alpha_1,\alpha_2}$. 
	\end{theorem}

	The thresholds $t^1_{OR}$ and $t^2_{OR}$ in the oracle rule \eqref{oracle-rule} can be approximately calculated using the following step-wise algorithm. Denote ${T}^{(i)}$ the $i$th ordered statistic of $\{{T}^1,..., {T}^m\}$. 
	{\small	\begin{equation}\label{alphacut-or}
			\mbox{Let}\, k_1= \min\left\{1\leq j\leq m: \frac{1}{j+1} \sum_{i=0}^j\left\{1-{T}^{(m-i)}\right\}\leq\alpha_1\right\},\, k_2= \max\left\{1\leq j\leq m: \frac 1 j \sum_{i=1}^j {T}^{(i)}\leq\alpha_2\right\}.
		\end{equation}}
As indicated by the theory in Section \ref{theory:sec}, $t^1_{OR}$ and $t^2_{OR}$ can be consistently estimated by $\hat t^1_{OR}=\min\left({T}^{(k_2)},0.5\right)$ and  $\hat t^2_{OR}=\min\left(1-{T}^{(m-k_1)},0.5\right)$ under mild conditions. Here 0.5 is imposed to avoid overlapping selections. To see why the stepwise algorithm \eqref{alphacut-or} makes sense, note that the moving average $\frac 1 r\sum_{j=1}^r {T}^{(j)}$ provides an estimate of $\mbox{mFSR}^2$ when $r$ observations with the smallest ${T}^j$ are selected to class 2 (cf. \cite{SunCai07}). Hence, it follows from \eqref{alphacut-or} that $\hat {t}^2_{OR}$ corresponds to the largest threshold such that the estimated $\mbox{FSR}^2$ is below $\alpha_2$. The explanation for $\hat t^1_{OR}$ is similar. 
	
	Denote $\pmb\delta^*_{OR}=\left\{\mathbb{I}(1-{T}^j<\hat t^1_{OR})+2\cdot\mathbb I({T}^j<\hat t^2_{OR}): 1\leq j\leq m\right\}$. The next theorem shows that the stepwise algorithm \eqref{alphacut-or} is valid for both FSR and mFSR control. 
	
	\begin{theorem} \label{thm:optimal_or2} 
		Consider the oracle setting where ${T}^j$ are known, $j=1, \cdots, m$. Then we have $\mbox{FSR}^k(\pmb\delta_{OR}^*)\leq\alpha_k$, $\mbox{mFSR}^k(\pmb\delta_{OR}^*)\leq\alpha_k$, for $k=1, 2$.
	\end{theorem}
	\begin{remark}
	    $\pmb\delta^*_{OR}$ is asymptotically optimal in the sense that $ECC_{\pmb{\delta}^*_{OR}}/ECC_{\pmb{\delta}_{OR}}\rightarrow 1$ as $m\rightarrow \infty$. This fact can be proved using similar arguments as those presented in the proof of Theorem \ref{thm3}.
	\end{remark}
	\subsection{Issues and roadmap} 
	
	The FSR control in selective classification, which is closely related to the false discovery rate (FDR, \citealp{BenHoc95}) control in multiple testing, presents unique challenges in high-dimensional inference. In multiple testing the null distribution of the $p$-values is assumed to be known precisely; hence FDR rules, such as the Benjamini-Hochberg's algorithm, can be derived to determine a proper $p$-value threshold that upper bounds the FDR. However, in classification the scores ($S_j^\pi$ or ${T}^j$) must be estimated from the training data with noise. For state-of-the-art LDA rules in the high-dimensional setting (\citealp{cai2011direct, shao2011sparse, Maietal12, dicker2016high, tony2019high}), the distributions of the estimated scores (and hence $p$-values) are in general unknown, rendering the uncertainty quantification and analysis of error rate intractable. 
	
	We take a different approach and develop a data-driven FSR rule in  two steps.  In the first step, we provide an efficient and robust score $\hat{S}_j$, which employs a new shrinkage rule that works well across sparse and dense regimes. In the second step, we develop a step-wise algorithm based on $\hat{S}_j$. Owing to the easy-to-analyze shrunken mechanism, we show that the uncertainties in the estimated score and its stochastic contribution to the errors by running the algorithm can be precisely quantified, establishing the theory on FSR control. 
	
	\setcounter{equation}{0}

	\section{The Data-Driven LASS Procedure}\label{estimation:sec}
	
	The key step in estimating the score $S^\pi_j$ is to develop a good estimate for $\pmb d=\pmb\mu_1-\pmb\mu_2$. In high-dimensional settings, most regularized LDA rules bet on the sparsity of $\pmb d$ (e.g. \citealp{Tibetal03, shao2011sparse}) to reduce the high variability in the sample estimates. However, the sparsity requirement, which may not hold in practice and often only serves as a working assumption, is  counter-intuitive in the sense that the two classes are better separated as $\pmb d$ has more nonzero elements. By contrast, \cite{efron2009empirical, greenshtein2009application, dicker2016high} proposed LDA rules based on Tweedie-type shrinkage estimators of $\pmb d$,  sidestepping the sparsity assumption. Existing non-sparse LDA rules have two limitations. First, Tweedie-type estimates are intractable to analyze, making it difficult to assess the uncertainties in classification. Moreover, Tweedie's formula requires that the elements in $\pmb d$ must be independent, which leads to efficiency loss when the dependence structure is highly informative \citep{cai2011direct, shao2011sparse}. We propose an easy-to-analyze shrinkage estimator that overcomes the above limitations. The methodology and illustrative examples are provided in Sections \ref{sec:method-shrink} and \ref{sec:example}. The data-driven LASS procedure is presented in Section \ref{sec:lass_proc}.  
	
	
	\subsection{Methodology}\label{sec:method-shrink}
	
	Let $\bar{X}_k$ and $\bar{Y}_k$ be the $k$th coordinate of $\bar{\X}$ and $\bar{\Y}$, respectively. We consider a class of shrinkage estimators
	\beq\label{cwse}
	\hat{\pmb{d}}=(\hat{d}_k: 1\leq k\leq p)=\left\{(\bar{X}_k-\bar{Y}_k){q}_k: 1\leq k\leq p \right\},
	\eeq
	where $q_k\in (0, 1)$ is a coordinate-wise shrinkage factor. To effectively reduce the uncertainties and quantify the associated misclassification risks, $q_k$ needs to be designed carefully such that it converges to $1$/$0$  at appropriate rates according to the strength of the signal. The proposed method chooses the following class of $q_k$:
	\begin{equation}\label{shrink:formal}
		q_k\coloneqq \dfrac{{g}_{1k}\left(|\bar{X}_k-\bar{Y}_k|\right)}{g_{0}\left(|\bar{X}_k-\bar{Y}_k|\right)+{g}_{1k}\left(|\bar{X}_k-\bar{Y}_k|\right)} ,
	\end{equation}
	where $g_{0}$ and ${g}_{1k}$ are respectively the density functions of $\mathcal{N} \left(0,\nn\right)$ and $$\mathcal{N}\left(\textcolor{black}{\left\{(2+b)\sqrt{\hat{\sigma}_{kk}}+\sqrt{(2+b)^2\hat{\sigma}_{kk}+4}\right\}}\sqrt{\ninv\log p},\nn\right),$$ $b>0$ is a small constant, and $\hat{\sigma}_{kk}$ is the pooled sample variance of {  $\{X_{ik}: i=1\ldots,n_1\}$ and $\{Y_{ik}: i=1\ldots,n_2\}$}.
	The constant $b>0$ is included in the definition only for theoretical considerations. In practice we can choose $b\approx 0$ or simply set $b=0$. In all our simulations and data analyses, we report the results with $b=0.1$, which are almost identical to the results with $b=0$. 
	
	The behavior of $q_k$ is qualitatively different depending on the strength of $d_k$. The following proposition shows an intuitively appealing demarcation phenomenon of $q_k$, implying that the multiplicative shrinkage rule \eqref{cwse} produces effects similar to that of hard thresholding rules: strong signals are kept and moderate/weak signals are suppressed. 
	
	\begin{proposition}\label{thm2}
		Consider ${q}_k$ defined in (\ref{shrink:formal}). Let $a_k={\left\{(2+b)\sqrt{\sigma_{kk}}+\sqrt{(2+b)^2\sigma_{kk}+4}\right\}}$ and $\epsilon$ an arbitrarily small constant. Define the following three groups:
		\begin{description}
			\item $\mathcal G_1=\left\{1\leq k\leq p: |d_k|> (a_k/2+\epsilon)\sqrt{\ninv\log p}\right\}$ (strong signals);
			\item $\mathcal G_2=\left\{1\leq k\leq p: |d_k|=o(\sqrt{\ninv\log p})\right \}$ (weak signals);
			\item $\mathcal G_3=\left\{1\leq k\leq p: |d_k|<(a_k/2-\epsilon)\sqrt{\ninv\log p} \mbox{ and } k\notin \mathcal G_2\right\}$ (moderate signals). 
		\end{description}
		Then there exists $\gamma>0$ independent of $p$, $n_1$ and $n_2$ such that the following hold:
		\begin{description}
			\item (a). $ 1- \mathbb{E} (q_k \mid k\in \mathcal G_1)=O(p^{-\gamma})$;
			\item (b). $ \mathbb{E} (q_k \mid k\in \mathcal G_3)=O(p^{-\gamma})$;
			\item (c). $ \mathbb{E} (q_k \mid k\in \mathcal G_2)=O(p^{-(1+\gamma)})$.
		\end{description}
	\end{proposition}

	We mention some merits of the proposed shrinkage rule. First, under the dense regime, the multiplicative factor $q_k$ can produce significantly less noisy estimates than original observations while being capable of retaining more nonzero coordinates than thresholding rules. This leads to shrinkage rules (Section \ref{sec:example}) with robust and superior performance at different sparsity levels. Second, unlike LDA rules based on Tweedie's formula \citep{efron2009empirical, dicker2016high}, the coordinate-wise shrinkage scheme in \eqref{shrink:formal} does not require the independence between $d_k$. Finally, the multiplicative rule is easy to analyze and leads to provably valid rules for FSR control.

	\subsection{Adaptation to unknown sparsity: illustrations}\label{sec:example}
	
	We present numerical examples to provide insights on why the shrinkage rule (\ref{cwse}) works well across the sparse and dense regimes. 
	
\begin{figure}
		\begin{center}
			\includegraphics[width=5.5in]{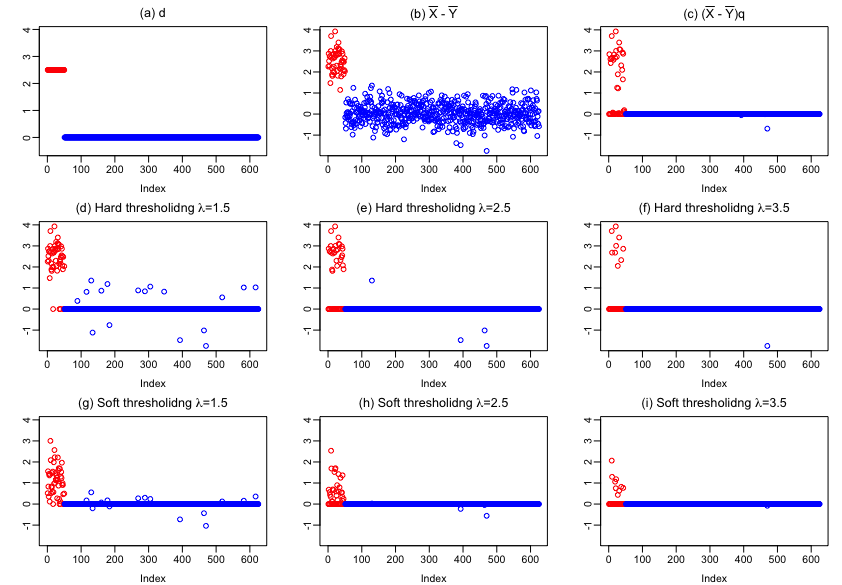} 
			\begin{center}
				\includegraphics[width=5.5in]{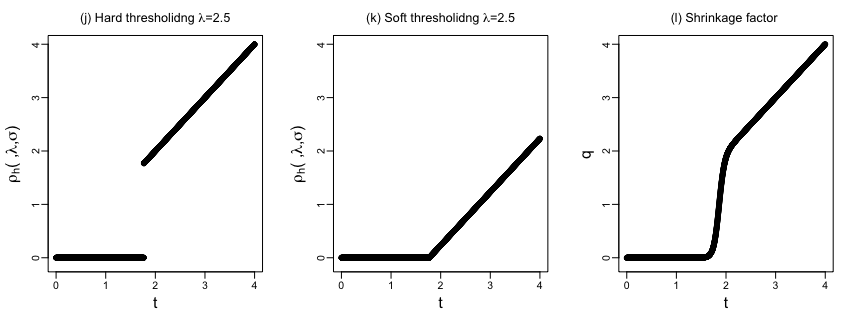} 	
			\end{center}
			\caption{\small A comparison of our shrinkage rule \eqref{shrink:formal} with hard/soft thresholding rules in the sparse case. Panels (a), (b) and (c) respectively plot the true ${d}_k$, observed $\bar{X}_k-\bar{Y}_k$ and $(\bar{X}_k-\bar{Y}_k)q_k$. Panels (d)-(f) [((g)-(i)] present results of hard-thresholding (soft-thresholding) rules. The effects of different shrinkage functions are provided at the bottom row.}\label{illusparse}
		\end{center}	
	\end{figure}

	We start with the sparse case. Suppose $n_1=n_2=n$. Let $n=5$, $p=625$, $\pmb{d}=(d_1, \ldots, d_p)$, where $d_k=2.5$ for $k\in \{1, \ldots, 50\}$ and $d_k=0$ for $k\in \{51, \ldots, 625\}$. The observations are generated as $\pmb{X}_i\sim \mathcal{N}(\pmb 0,\frac 1 2 I_p)$ and $\pmb{Y}_i\sim \mathcal{N}(\pmb{d}, \frac 1 2 I_p)$, $i=1, \ldots, n$, where $\pmb 0$ is a $p$-dimensional vector of zeros and $I_p$ a $p$-dimensional identity matrix. It follows that $\bar{X}_k-\bar{Y}_k\sim \mathcal{N}(d_k,1/n)$. We contrast the proposed shrinkage rule with hard/soft thresholding rules in Figure \ref{illusparse}. Panels (a) and (b) plot $d_k$ and observed $\bar{X}_k-\bar{Y}_k$, respectively, with coordinates corresponding to zero/nonzero $d_k$ being marked in  blue/red. Panel (c) plots the shrinkage estimates $(\bar{X}_k-\bar{Y}_k)q_k$, where $q_k$ is defined in (\ref{shrink:formal}) with $b=0$. Panels (d)--(f) show the hard/soft thresholding estimates with different thresholds $\lambda$, where $\rho_h(t,\lambda,\sigma)\coloneqq t\mathbb{I}(|t|>\lambda\sqrt{\sigma})$ and $\rho_s(t,\lambda,\sigma)\coloneqq \sgn{(t)}\cdot\max(|t|-\lambda\sqrt{\sigma},0)$ are the hard and soft thresholding functions, respectively.
	
	
	We can see from Panel (c) that almost all blue points are pulled towards and centered around the line of zero by our proposed shrinkage rule. The multiplicative factor $q_k$ is as effective as existing thresholding methods for noise reduction in the sense that the patterns in Panel (c) is qualitatively similar to those in Panels (d) to (f), except that in (d) and (e) too many blue points have survived whereas in (f) too many nonzero signals have been killed. Panel (c) shrinks most noisy entries to zero while being capable of preserving a significant portion of nonzero signals. The bottom row compares the shrinkage functions $\rho_h$ and $\rho_s$ with $q_k$ by setting $\hat{\sigma}_{kk}=0.5$ and $\lambda=2.5$. We can see that our shrinkage function is continuous, nearly unbiased for large signals and yields similar effects as that of the hard-thresholding function. The proposed shrinkage rule is desirably tuning-free in the sense that one can simply set $b=0$ in practice; this merit is justified in our theoretical analysis and corroborated by our numerical results. By contrast, the performance of existing thresholding rules depends critically on the value of $\lambda$, which is nontrivial to choose.

\begin{figure}
		\begin{center}
			\includegraphics[width=5.5in, height=3.2in]{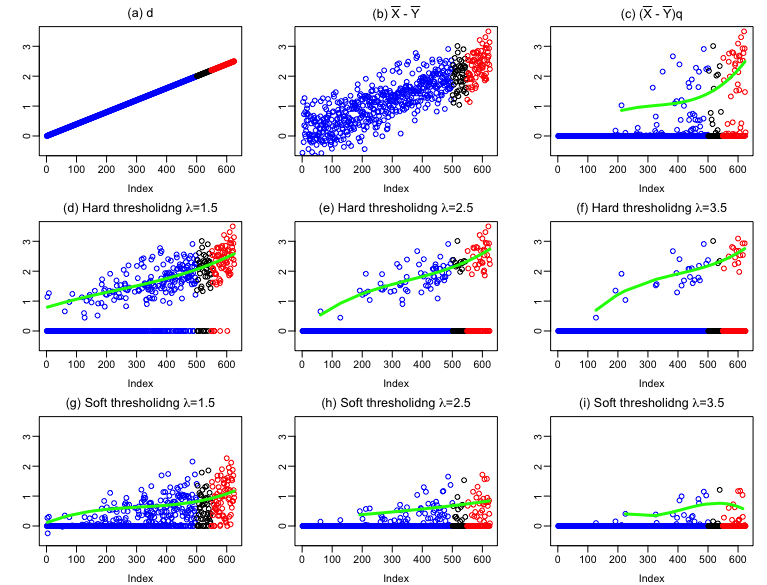} 
			\caption{Comparison of the proposed shrinkage rule with thresholding rules.  The proposed shrinkage rule exhibits a curved pattern [Panel (c)], which is more effective in eliminating weak signals and keeping strong signals. Green lines are cubic polynomial fits of the points above the horizontal line of $0.2$. }
			\label{illus}
		\end{center}	
	\end{figure}

	Next we turn to the dense case. The data are generated in a similar way as before except that we let $d_k=2.5(k-1)/624$, $k=1, \ldots, 625$, i.e. $d_k$ increases linearly from 0 to 2.5. We mark the coordinates of $\pmb d$ in three colors: red if $d_k>(\sqrt{0.5}+\sqrt{0.5+1})\sqrt{\log p/n}\approx2.19$, black if $1.5<d_k<(\sqrt{0.5}+\sqrt{0.5+1})\sqrt{\log p/n}$ and blue otherwise\footnote{There are no clearcut boundaries; the colors in this example are only chosen for illustration purpose.}. We plot our shrinkage estimate and thresholding estimates in Figure \ref{illus}. In this high-dimensional ``dense'' regime, the boundaries between weak, moderate and strong signals are blurred. Therefore the working assumptions (sparsity and dichotomy of $\pmb d$) underpinning the use of thresholding rules can be problematic. In contrast with the (roughly) linear patterns of the surviving points in Panels (d) to (i), the curved pattern in Panel (c) provides differential amount of shrinkage for weak and strong signals, achieving a more desirable balance between reducing the uncertainties and maintaining useful signals. 
	
	
	\subsection{The Data-Driven LASS Procedure}\label{sec:lass_proc}
	
	We propose to estimate $S^\pi_j=(\pmb W_j-\pmb\mu)^\top\Sigma^{-1}\pmb d$ by
	\begin{equation}\label{shat}
		\hat{S}_j=\left(\pmb{W}_j-\dfrac{\bar{\pmb{X}}+\bar{\pmb{Y}}}{2} \right)^\top\hat{\Sigma}^{-1}\hat{\pmb{d}}.
	\end{equation}
	First, $\hat{\Sigma}^{-1}$ is the estimated precision matrix and $\hat{\pmb{d}}$ is the proposed shrinkage estimate \eqref{cwse}. The estimation of the precision matrix has been intensively studied in the literature; see \cite{liu2015fast}, \cite{cai2016estimating}, \cite{loh2018high}, \cite{wang2013large}, \cite{sun2013sparse}, \cite{yuan2010high} for related works. In our numerical studies we use the ACLIME estimator proposed in \cite{cai2011constrained}. Next, $\hat{\pmb{d}}=((\bar{X}_1-\bar{Y}_1)q_1,\ldots,(\bar{X}_p-\bar{Y}_p)q_p )$, where $q_k$ is as defined in (\ref{shrink:formal}). Let
	$
	\hat {T}^j:=\dfrac{\exp(\hat{S}_j) }{1+\exp(\hat{S}_j)}.
	$
	Denote $\{\hat{T}^{(j)}: 1\leq j\leq m\}$ the ordered statistics. Define
	{\small
		\begin{equation}\label{alphacut}
			k_1= \min\left\{1\leq j\leq m: \frac{1}{j+1} \sum_{i=0}^j (1-\hat{T}^{(m-i)})\leq\alpha_1\right\},\quad 	k_2= \max\left\{1\leq j\leq m: \frac 1 j \sum_{i=1}^j \hat{T}^{(i)}\leq\alpha_2\right\}.
	\end{equation}}
	The data-driven LASS procedure is given by $\hat{\pmb\delta}=(\hat{\delta}_1,\ldots \hat{\delta}_m)$, where
	\begin{equation}\label{de}
		\hat{\delta}_j=\mathbb{I}\left\{ 1-\hat{T}^j<  \min\left(1-\hat{T}^{(m-k_1)},0.5\right)\right\}+2\cdot\mathbb{I}\left\{ \hat{T}^j\leq \min\left(\hat{T}^{(k_2)},0.5\right)\right\}.
	\end{equation}
	
	\begin{remark}\rm{
			If we choose $\alpha_1=\alpha_2=0.5$, then indecisions are not allowed by \eqref{de}. That is, LASS becomes a classical rule that makes definitive classifications on all individuals. We shall see that LASS is still superior in both theory and numerical performance compared to existing methods under this classical setup (Corollary \ref{coro1} in Section \ref{theory:sec} and Section \ref{simu:conv}). }
	\end{remark}

	\setcounter{equation}{0}
	
	\section{Theoretical Properties of LASS} \label{theory:sec}
	
	This section studies theoretical properties for the data-driven LASS procedure. We focus on the regime of $\nn\log p \rightarrow 0$, which requires that the dimension does not grow too fast relative to the sample size. We consider issues on FSR control and optimality in turn, and conclude the section with a discussion of connections to existing works. 
	
	We first state and explain a few conditions needed in our theoretical analysis. 

	\noindent\textbf{(A1)} The covariance matrix $\Sigma=(\sigma_{kl})_{1\le k, l\le p}$ satisfies $0<\epsilon_0\le \sigma_{kk}\le1/\epsilon_0$ for all $1\le k\le p$, where $\epsilon_0$ is a fixed positive constant.
	
	(A1) is a standard condition in matrix analysis, and is satisfied when the covariance matrix is \emph{well-conditioned}, as assumed in, for example, \cite{bickel2008regularized}. 
	
	\noindent \textbf{(A2)} The estimated precision matrix $\hat{\Sigma}^{-1}$ satisfies $\|\hat{\Sigma}^{-1}- \Sigma^{-1} \|^2_2=o(1) $.
	
	Consistent estimation of the precision matrix $\Sigma^{-1}$ has been studied intensively. Effective estimators and sufficient conditions for consistent estimation have been discussed in  a large body of works including \cite{bickel2008regularized}, \cite{yuan2010high}, \cite{liu2015fast},\cite{cai2016estimating} and \cite{avella2018robust}, among others.
	
	\noindent\textbf{(A3)} $|\mathcal{G}_1|\geq 1$ and $|\mathcal G_3|=O(\frac{n_1n_2}{n_1+n_2})$, where $\mathcal G_1$ and $\mathcal G_3$ are defined in Proposition \ref{thm2} and correspond to collections of strong and moderate coordinates of $\pmb d$, respectively. 
	
	In (A3), $|\mathcal{G}_1|\geq 1$ provides a sufficient condition under which LASS makes at least one definitive classification with high probability. As opposed to existing works that require the sparsity of both $\mathcal G_1$ and $\mathcal G_3$, we do not impose an upper bound on $|\mathcal G_1|$. Our condition seems to be more sensible because having more strong signals ($d_k\in\mathcal G_1$) is helpful for distinguishing the two classes, and what really hurts the performance of LDA rules is an overwhelming number of nonzero elements of moderate strength ($d_k\in\mathcal G_3$). The condition $|\mathcal G_3|=O(\ntt)$ corresponds to a weaker notion of sparsity in the sense that the sparsity or approximate sparsity conditions in existing works (e.g. \citealp{cai2011direct}) will be violated if $|\mathcal G_3|\gg \ntt $. We stress that the conventional sparsity notion assumes that there are relatively few signals, while we require that there are relatively few signals \emph{of moderate strength} will fall with the narrow range defined by $\mathcal G_3$, which eliminates the need of the counter-intuitive sparsity condition on $\mathcal G_1$ as used in existing works. The superiority of LASS under the dense signal setting is illustrated in our numerical results (Section \ref{simu}).

	The next theorem establishes the asymptotic validity of LASS for FSR control. 
	
	\begin{theorem}\label{thm4} Let $\hat{\pmb{\delta}}$ be the data-driven LASS procedure  defined in \eqref{de}. Under conditions (A1)-(A3), we have
		$\mbox{mFSR}^c_{\hat{\pmb{\delta}}}\leq \alpha_c+o(1)$ and $\mbox{FSR}^c_{\hat{\pmb{\delta}}}\leq \alpha_c+o(1)$ for $c=1,2$. 
	\end{theorem}
	\begin{remark}
	Note that in Theorem \ref{thm4} $\mbox{mFSR}^c_{\hat{\pmb{\delta}}}$ can be asymptotically smaller than $\alpha_c$. This is because when the two classes are well separated, $\hat{T}^{(k_2)}$ and $1-\hat{T}^{(m-k_1)}$  can be greater than 0.5 ($k_1$ and $k_2$ are as defined in \eqref{alphacut}). 	Since the thresholds in \eqref{de} are at most 0.5, $\mbox{mFSR}^c_{\hat{\pmb{\delta}}}$ may not attain the target level $\alpha_c$. However, this should not be viewed as a drawback of our method. 
	If both $\hat{T}^{(k_2)}$ and $1-\hat{T}^{(m-k_1)}$  are larger than 0.5, there will be no indecision and we are back in the classical setting. 
\end{remark}
	Our conditions on error rate control are substantially different in nature compared to those required by state-of-the-art LDA rules. First, the sparsity of $\Sigma^{-1}$ is not a necessary condition for our theory on FSR control\footnote{Note that even when the sparsity of $\Sigma^{-1}$ is needed for consistent estimation, the sparsity conditions on $\Sigma^{-1}$ and $\pmb{d}$ usually correspond to fundamentally different notions in scientific studies. Our theory seems to be more sensible as it eliminates the need on the sparsity of $\pmb d$.}. Second, our theory needs neither sparsity nor consistent estimation of $\pmb d$. Finally, in contrast with existing works (e.g. \cite{cai2011direct}, \cite{shao2011sparse}), our theory has no restrictions on the norm of $\pmb{d}$ or $\Sigma^{-1}\pmb{d}$. We stress that estimation and classification are fundamentally different tasks: the assumptions on the norm are natural for estimation problems, but counter-intuitive for classification problems -- larger norms make the classification task easier and lead to lower error rates. 
	
	Next we investigate the asymptotic optimality of LASS. The next condition, which requires the elements in both $\mathcal G_2$ and $\mathcal G_3$ are sparse in $L_2$, ensures that shrinking weak and moderate signals has negligible effects on the power. 
	
	\noindent \textbf{(A4)} $\sum_{k\notin \mathcal{G}_1}d^2_k=o(1)$.
	
	The next theorem shows that the data-driven LASS achieves the performance of the oracle procedure \eqref{oracle-rule} asymptotically. As opposed to existing theories in sparse LDA, no conditions are imposed on the sparsity or norm of strong signals ($d_k\in \mathcal G_1$). 
	
\begin{theorem}\label{thm3}
	Under conditions (A1)-(A4), we have $\mbox{mFSR}^c_{\hat{\pmb{\delta}}}\leq\alpha_c+o(1)$, $\mbox{FSR}^c_{\hat{\pmb{\delta}}}\leq\alpha_c+o(1)$ for $c=1,2$, and {$ECC_{\hat{\pmb{\delta}} }/ECC_{ \pmb{\delta}_{OR}}=1+o(1)$.} 
\end{theorem}

	If we let $\alpha_1=\alpha_2=0.5$, then the FSR control setup reduces to the classical setup where indecisions are not allowed. Let $\pmb{\delta}$ be a classification rule that only takes on value 1 or 2, define $L(\pmb{\delta})=\mathbb{P}\left\{\theta_j\neq \delta_j| (\pmb{X}_i, 1\leq i\leq n_1), (\pmb{Y}_i, 1\leq i\leq n_2)\right\}$, $R(\pmb{\delta})=\mathbb{E}\left\{ L(\pmb{\delta}) \right\}$,
	the following corollary is a direct consequence of Theorem \ref{thm3}.
	\begin{corollary}\label{coro1} (Risk consistency).
		Suppose we choose $\alpha_1=\alpha_2=0.5$. Then under conditions (A1)-(A5), we have $R(\hat{\pmb{\delta}})-R(\pmb \delta^F)\rightarrow 0$, 
		where $\pmb \delta^F$ is the oracle Fisher's rule.
	\end{corollary}
	
	We conclude this section with an example to contrast our thoery with existing theories in high-dimensional LDA, where the assumption on the sparsity of $\pmb d$ is commonly believed to be ``indispensable''. For example, \cite{cai2011direct} requires  $|\Sigma^{-1}\pmb{d}|_0=o(\sqrt{(n_1+n_2)/\log p})$ to achieve risk consistency. 
	This sparsity condition seems to be necessary if the scope of analysis is limited to the class of thresholding rules. It is not an artifact of the theoretical analysis, as we can see from the numerical results in Section \ref{simu:dense} where the disadanvantages of existing works are reflected.
However, LASS still achieves risk consistency even when the condition in \cite{cai2011direct} is violated. We illustrate this through the next example.
	
	\begin{example}\rm{
			Consider an asymptotic setup where $n_1=n_2=n,\ p=n^2$, $\Sigma=I_p$ and $\pmb{\mu}_1=(0,\ldots,0)^\top$. Let $\pmb{\mu}_2=(1,\ldots,1,0,\ldots,0)^\top$ be a vector with the first $k$ entries being $1$ and the rest 0. Let $k=n$ and $\Delta_p=\pmb{d}^\top\pmb{\Sigma}^{-1}\pmb{d}$, then in this setting $\Delta_p=n$. It is clear that $|\pmb{\Sigma}^{-1}\pmb{d}|_0=n$ and $|\pmb{\Sigma}^{-1}\pmb{d}|_1=n$. Denote $\hat{\pmb{\delta}}_{LPD}$ the LPD rule in \cite{cai2011direct}. To guarantee that $R(\hat{\pmb{\delta}}_{LPD})-R(\pmb{\delta}^F)\rightarrow 0$, the theory in \cite{cai2011direct} requires that $|\pmb{\Sigma}^{-1}\pmb{d}|_0=o(\sqrt{n/\log p})$ or $\frac{|\pmb{\Sigma}^{-1}\pmb{d}|_1}{\Delta_p^{1/2}}+\frac{|\pmb{\Sigma}^{-1}\pmb{d}|^2_1}{\Delta_p^{2}}=o(\sqrt{n/\log p})$, both of which are violated. By contrast, note that all nonzero elements in $\pmb d$ are in $\mathcal G_1$, the conditions of Corollary \ref{coro1} are satisfied for any $k\geq 1$, which guarantees that LASS achieves risk consistency across sparse and dense settings. We shall see in simulation studies that the gap in numerical performance between LASS and LPD grows as $k$ increases.} 
	\end{example}
	
As opposed to existing works that produce approximately the same amount of shrinkage for elements in both $\mathcal G_1$ (strong) and $\mathcal G_3$ (moderate), LASS adopts an adaptive strategy that makes it possible to provide differential amounts of shrinkage for the elements in $\mathcal G_1$ and $G_3$ (Proposition 1 and Figure 2c). The resulting shrinkage rule is more effective in keeping strong signals and eliminating noisy elements; this explains the superiority of LASS in both theory and numerical performance.
	
	\section{Numerical Experiments}\label{simu}		
	
	This section illustrates the numerical performance of LASS using both simulated and real data. The simulation considers two setups: the conventional setup that does not allow indecisions (Section \ref{simu:conv}), and the selective classification setup that aims to control the FSR (Section \ref{simu:FSR}). Two real data sets are discussed in Sections \ref{data:sec1} and \ref{data:sec2}. 
	In all analyses, LASS is implemented using $b=0.1$ in (\ref{shrink:formal}) and the ACLIME method \citep{cai2016estimating} is adopted for estimating $\Sigma^{-1}$. For the simulated data, we take $n_1=n_2=n$.

	\subsection{Simulation: the conventional setup}\label{simu:conv}
	
	We start with the classical setting where no indecisions are allowed. We compare LASS with the following methods.
	\begin{itemize}
		\item Fisher's rule that uses true $\pmb{\mu}_1$, $\pmb{\mu}_2$ and $\Sigma^{-1}$ (denoted ``Oracle''). This method serves as the optimal benchmark for all classification rules.
		\item The LPD rule proposed by \cite{cai2011direct} (denoted ``LPD''), which is implemented using the code provided on the authors' website. 
		\item The AdaLDA rule proposed by \cite{tony2019high} (denoted ``AdaLDA''), which is implemented using the code provided on the authors' website. 
		\item Fisher's rule that uses sample estimates of $\pmb{\mu}_1$, $\pmb{\mu}_2$ and $\Sigma^{-1}$ (denoted ``Naive''). Specifically, $S^\pi_j$ is estimated as $\left(\pmb{W}_j-\frac{{\bar{\pmb{X}}+\bar{\pmb{Y}}}}{2}  \right)^\top\hat{\Sigma}^{-1}(\bar{\pmb{X}}-\bar{\pmb{Y}})$, where $\hat{\Sigma}^{-1}$ is the Penrose inverse of the sample covariance matrix. 
		\item The $L_1$ logistic regression method (denoted ``Lasso''). We have followed the approach suggested by \cite{lei2014classification}, where the tuning parameter is chosen by cross-validation.
		
		\item The empirical Bayes method proposed in \cite{efron2009empirical} (denoted ``Ebay''). We use the R-package \texttt{Ebay} to estimate $\pmb{d}$. $\Sigma^{-1}$ is estimated by a diagonal matrix where diagonals are the inverses of sample variances. 
	\end{itemize}
	
	We present numerical results in the next two subsections to show that LASS (a) is comparable to state-of-the-art methods in the sparse case, and (b) substantially outperforms competing methods in the dense case.

	\subsubsection{Sparse setting}\label{simu:sparse}
	
	Let $\pmb{\mu}_1=(0,...,0)^\top\in \mathbb{R}^p$, and $\pmb{\mu}_2$ be a vector with the first $10$ entries being 0.5, the next 10 being $0.1(\log p/n)^{1/2}$, and the rest being 0. Consider the following three correlation structures that are widely considered in the literature \citep{cai2011direct, tony2019high, avella2018robust}. 
	
	\begin{description}
		\item \emph{Model 1: Band graph.} Let $\Sigma^{-1}=\Omega=(\omega_{ij})_{p\times p}$, where $\omega_{ii}=1$, $\omega_{i,i+1}=\omega_{i+1,i}=0.35$, $\omega_{i,i+2}=\omega_{i+2,i}=0.175$, and $\omega_{ij}=0$ if $|i-j|>2$. 
		
		\item \emph{Model 2: AR(1) structure}. Let $\Sigma^{-1}=\Omega=(\omega_{ij})_{p\times p}$, where $\omega_{ij}=0.3^{|i-j|}$.

		\item \emph{Model 3: Block structure}. Let $\Sigma^{-1}=\Omega=(\mathbf{B}+\delta I_p)/(1+\delta)$, where $b_{ij}=b_{ji}=0.05\cdot \text{Bernoulli}(0.1)$ for $1 \leq i \leq p/2, i < j \leq p$, $b_{ij} = b_{ji} = 0.05$ for $p/2 + 1 \leq i < j \leq p$, $b_{ii} = 1$ for $1 \leq i \leq p$, and $\delta=\max\{-\lambda_{min}(\mathbf{B}),0\}+0.1$.
		
	\end{description}
	
	The size of the training set is $n=400$, with $p$ varying from 500 to 1000. The mis-classification rate is computed based on $m=2000$ test points generated from $\mathcal{N}(\pmb{\mu}_1,\Sigma)$ or $\mathcal{N}(\pmb{\mu}_2,\Sigma)$ with equal probability. We repeat the experiment for 100 times, and report the misclassification rates (in percentage) in Table \ref{table1}. 
	
	\makeatletter\def\@captype{table}\makeatother 
	\begin{center}
		\begin{spacing}{1}
			
			\begin{tabular}{l c c c c c c r} 
				$p$&\textbf{Oracle}&	\textbf{Naive} &\textbf{LASS}&  \textbf{LPD} & \textbf{AdaLDA}& \textbf{LASSO}& \textbf{Ebay}\\
				\hline
				\hline
				\multicolumn{8}{c}{Model 1}\\
				\hline
				
				500 & 13.74& 29.39& $\mathbf{14.78}$& 15.78& 14.79& 15.75& 15.93\\
				600 & 13.64& 33.12& 14.72& 15.52& $\mathbf{14.55}$& 15.78& 15.66\\
				700 & 13.81& 38.51& 15.02& 15.80& $\mathbf{14.85}$& 16.10& 15.99\\
				800 & 13.64& 47.72& 14.87& 15.89& $\mathbf{14.81}$& 16.12& 15.62\\
				900 & 13.70& 41.66& $\mathbf{14.44}$& 17.41& 14.75& 16.31& 15.79\\
				1000 & 13.70& 41.16& $\mathbf{14.59}$& 18.06& 14.73& 16.27& 15.92\\
				\hline
				\multicolumn{8}{c}{Model 2}\\
				\hline
				500 & 14.82& 30.68& 15.98& 16.61& $\mathbf{15.77}$& 16.72& 16.03\\
				600 & 14.81& 34.45& 16.15& 16.68& $\mathbf{15.77}$& 16.84& 16.35\\
				700 & 14.79& 39.08& 16.21& 16.67& $\mathbf{15.86}$& 16.92& 16.36\\
				800 & 14.71& 47.83& 16.13& 16.77& $\mathbf{15.91}$& 16.98& 16.02\\
				900 & 14.87& 41.79& 16.19& 18.14& $\mathbf{15.86}$& 17.08& 16.37\\
				1000 & 14.92& 41.25& 16.38& 18.72& $\mathbf{15.92}$& 17.11& 16.51\\
				\hline
				\multicolumn{8}{c}{Model 3}\\
				\hline
				500 & 21.16& 36.20& $\mathbf{22.93}$& 23.83& 23.71& 24.21& 23.31\\
				600 & 20.87& 39.44& $\mathbf{23.14}$& 24.23& 24.04& 24.69& 23.48\\
				700 & 21.00& 42.81& $\mathbf{23.52}$& 24.69& 24.49& 25.03& 23.88\\
				800 & 20.99& 48.59& $\mathbf{23.82}$& 25.00& 24.87& 25.28& 24.08\\
				900 & 21.02& 44.22& $\mathbf{24.29}$& 25.78& 25.37& 26.01& 24.48\\
				1000 & 21.05& 43.02& $\mathbf{24.72}$& 27.04& 26.11& 26.41& 25.08\\
				\hline
			\end{tabular}
			\caption{\label{table1} Comparison of average misclassification rate in percentage. The smallest error rate (next to that by the oracle) in each setting has been boldfaced. }
		\end{spacing}
	\end{center}

	We can see that the Naive method can be substantially improved by LPD, AdaLDA and LASSO, all of which make strong assumptions on the sparsity structure of the data generating model. Although no method dominates, LASS and AdaLDA seem to perform the best among all methods being considered. LASS is comparable to AdaLDA in terms of the overall effectiveness across the three settings. This is impressive since \cite{tony2019high} showed AdaLDA is minimax optimal in sparse LDA. We shall see in the next simulation that in the non-sparse setting, LASS substantially outperforms AdaLDA. Similar to LASS, the Ebay method adopts the shrinkage idea and does not make strong assumptions on the sparsity structure. We can see that Ebay performs reasonably well. However, Ebay relies on the independence assumption, has no theoretical guarantee on the convergence of error rate, and is less effective than LASS in all settings.  
	
	\subsubsection{Dense setting}\label{simu:dense}
	
	Consider the three models in the previous section. The choices of $\pmb\mu_1$ and $\Sigma$ are the same, while $\pmb\mu_2$ contains more nonzero entries: the first $(p/4)$ entries are 0.4 and the rest are 0. The misclassification rates (in percentage) are summarized in Table \ref{table2}. As expected, methods that rely heavily on the sparsity assumption of $\pmb d$, such as LPD and AdaLDA, do not perform well. We mention a few important patterns in the results. 
	\begin{itemize}
		\item The performance of LPD and AdaLDA deteriorates as $p$ increases. This is undesirable considering that the classification problem seems to have become easier, as is manifested in the improved performance of the oracle  rule. In many settings LPD and AdaLDA can be worse than Naive.  
		\item Ebay does relatively well when $p$ is small but its performance also deteriorates as $p$ increases. The misclassification rates can be much higher than the oracle benchmark.
		\item LASS and Lasso substantially outperform competing methods in most settings. The performance of both improves as $p$ increases, exhibiting the same desirable trend as that of the oracle rule. LASS dominates Lasso, and the gap in error rate is substantial in several settings. 
		
	\end{itemize}

	\subsection{Simulation: FSR control}\label{simu:FSR}
	
	We now turn to the selective inference setup where the goal is to control the FSR. 
	For Naive, Lasso and Ebay, we first form an estimate for the discriminant, denoted as $\hat{S}_j$, then use $\hat{T}^j=\frac{\exp(\hat{S}_j)}{1+\exp(\hat{S}_j)}$ in \eqref{alphacut} and \eqref{de}, which serve as the \emph{base algorithm} for FSR control. LPD and AdaLDA are omitted as they only produce the signs of the discriminants and it is unclear how to adjust the algorithms for FSR control. 
	
	Next we present results pertaining to FSR control in the sparse settings as considered in Section \ref{simu:sparse}, but omit the results for the dense settings in Sections \ref{simu:dense}. The reason is that when the classification task becomes easy (as indicated by Table \ref{table2}): the misclassification rate is so low that the FSR framework is no longer needed.   
	
	\makeatletter\def\@captype{table}\makeatother 
	\begin{center}
		\begin{spacing}{1}
			
			\begin{tabular}{l c c c c c c r} 
				$p$&\textbf{Oracle}& \textbf{Naive}& \textbf{LASS} & \textbf{LPD} & \textbf{AdaLDA}&  \textbf{Lasso}& \textbf{Ebay}\\
				\hline
				\hline
				\multicolumn{8}{c}{Model 1}\\
				\hline
				500 & 0.07 & 2.95 & 0.20 & 5.19 & 2.09 & 0.52 & $\mathbf{0.12}$\\
				600 & 0.02 & 4.48 & $\mathbf{0.09}$ & 5.00 & 3.18 & 0.31 & 0.42\\
				700 & 0.01 & 9.79 & $\mathbf{0.04}$ & 11.59 & 4.80 & 0.20 & 0.78\\
				800 & 0.00 & 39.79 & $\mathbf{0.03}$ & 15.24 & 6.38 & 0.16 & 1.09\\
				900 & 0.00 & 12.37 & $\mathbf{0.01}$ & 15.45 & 8.59 & 0.15 & 1.30\\
				1000 & 0.00 & 8.42 & $\mathbf{0.01}$ & 11.29 & 10.17 & 0.12 & 1.54\\
				\hline
				\multicolumn{8}{c}{Model 2}\\
				\hline
				500 & 0.08 & 3.02 & 0.23 & 5.48 & 2.05 & 0.54 & $\mathbf{0.16}$\\
				600 & 0.03 & 4.56 & $\mathbf{0.10}$ & 4.93 & 2.81 & 0.32 & 0.46\\
				700 & 0.01 & 10.08 & $\mathbf{0.05}$ & 11.80 & 4.38 & 0.23 & 0.79\\
				800 & 0.00 & 39.94 & $\mathbf{0.02}$ & 12.69 & 6.20 & 0.15 & 1.15\\
				900 & 0.00 & 12.67 & $\mathbf{0.01}$ & 15.27 & 8.14 & 0.13 & 1.32\\
				1000 & 0.00 & 8.36 & $\mathbf{0.01}$ & 11.31 & 9.42 & 0.10 & 1.56\\
				\hline
				\multicolumn{8}{c}{Model 3}\\
				\hline
				500 & 0.26 & 5.06 & $\mathbf{1.32}$ & 6.52 & 3.37 & 2.04 & 1.50\\
				600 & 0.08 & 6.35 & $\mathbf{0.86}$ & 6.22 & 3.78 & 1.37 & 1.69\\
				700 & 0.02 & 11.72 & $\mathbf{0.62}$ & 5.34 & 4.64 & 0.99 & 1.92\\
				800 & 0.01 & 39.89 & $\mathbf{0.44}$ & 5.32 & 6.36 & 0.70 & 2.05\\
				900 & 0.00 & 14.08 & $\mathbf{0.38}$ & 16.00 & 8.25 & 0.61 & 2.18\\
				1000 & 0.00 & 10.05 & $\mathbf{0.36}$ & 18.96 & 10.47 & 0.51 & 2.22\\
				\hline
			\end{tabular}
			\caption{\label{table2} Comparison of average misclassification rate in percentage. The smallest error rate (next to that by the oracle) in each setting has been boldfaced. }
		\end{spacing}
	\end{center}

\begin{figure}
		\begin{center}
			\includegraphics[width=6.5in, height=1.2in]{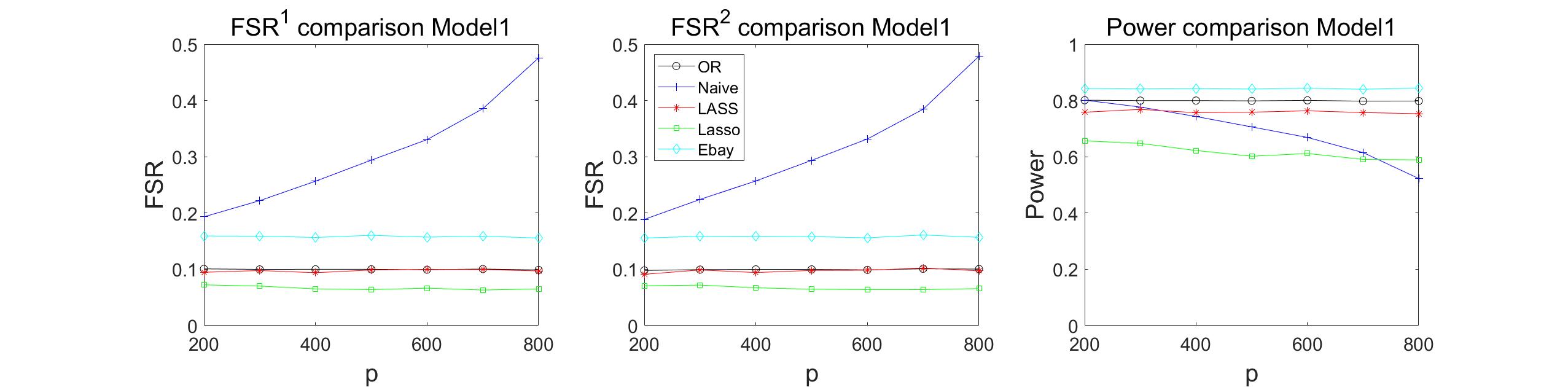}
			\includegraphics[width=6.5in, height=1.2in]{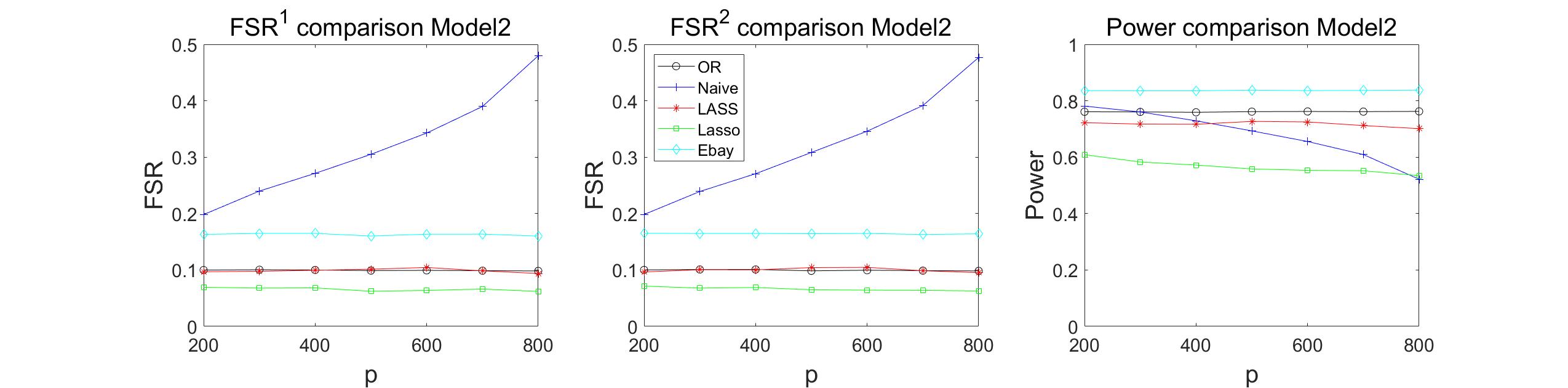}
			\includegraphics[width=6.5in, height=1.2in]{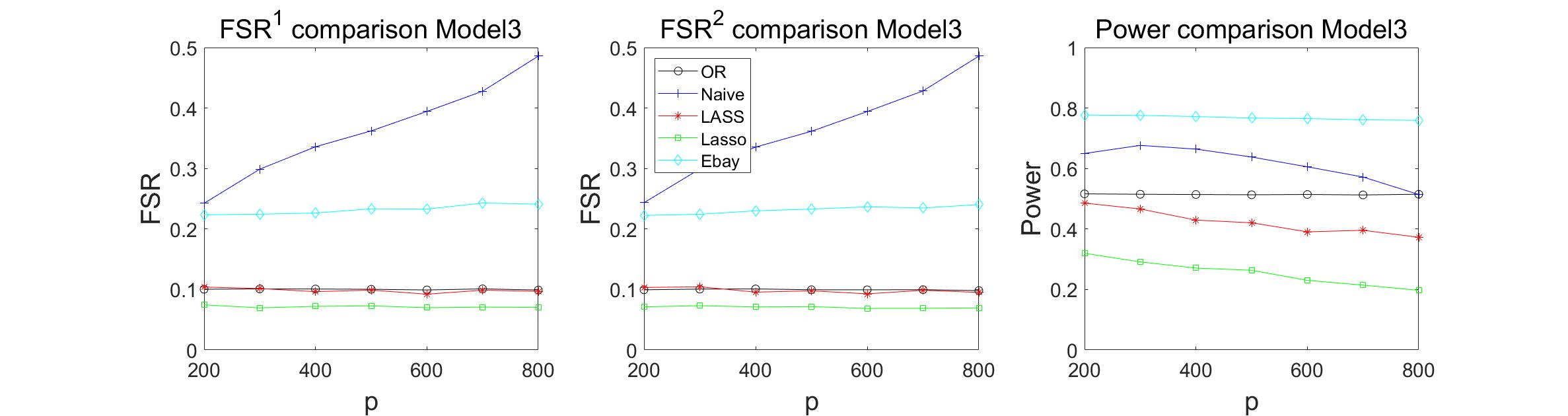} 
			\caption{\small Comparison of FSR and Power. Naive and Ebay fail to control the FSR. LASS controls FSR at the nominal level with the highest Power.}
			\label{plot:1}
		\end{center}
	\end{figure}

	Consider the models in Section \ref{simu:sparse}. We fix $n=400$ and vary $p$ from 200 to 800. The target $\text{FSR}^1$ and  $\text{FSR}^2$ levels are both set 0.1. The experiment is repeated for 100 times, and the average FSRs (displayed in the first two columns) and power (defined as ECC$/m$; displayed in the last column) are reported in Figure \ref{plot:1}. We mention the following patterns.

	\begin{itemize}
		\item Both Naive and Ebay fail to control the FSR. The Naive method becomes worse as $p$ increases. This corroborates the analysis in \cite{bickel2004some}, which shows that LDA rules based on sample estimates suffer from high dimensionality. 
		\item Both Lasso and LASS control the FSR at the nominal level, showing that our proposed data-driven algorithm \eqref{alphacut} is effective for FSR control when equipped with reasonably good estimates of the scores.  
		\item LASS controls the FSR at the nominal level accurately across all settings. Lasso is conservative and has lower power.  
		
	\end{itemize}

	\subsection{Lung cancer data}\label{data:sec1}
	
	We analyze the lung cancer data (\citealp{gordon2002translation}, available at {\small \url{https://leo.ugr.es/elvira/DBCRepository/LungCancer/LungCancer-Harvard2.html}}), a benchmark data set in high-dimensional classification problems. The data set collects expression levels for $p=12,533$ genes on $181$ tissue samples. Among the 181 samples, 31 are from malignant pleural mesothelioma (MPM) group and 150 are from adenocarcinoma (ADCA) group.  The training set contains 16 samples from MPM and 16 samples from ADCA. The testing set contains 134 samples from MPM and 15 samples from ADCA.
	
\begin{figure}
		\begin{center}
			\includegraphics[width=5.5in, height=2.8in]{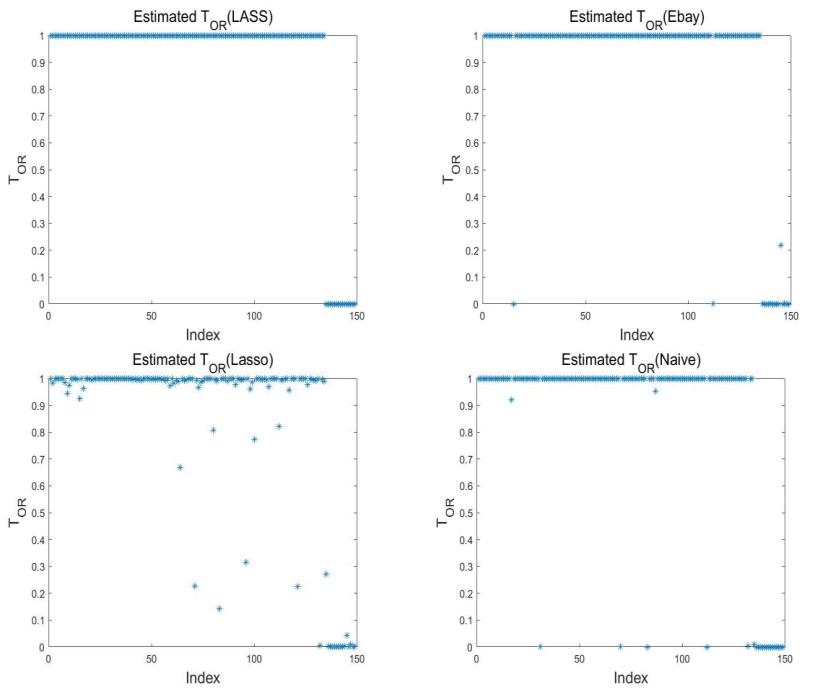} 
			\caption{Comparison of estimated $T$: LASS separates the two classes almost perfectly. Ebay does better than Naive and Lasso but worse than LASS.
			}\label{fig2}
		\end{center}
	\end{figure}

	We follow the same data pre-processing steps in \cite{cai2011direct}. First, the sample variances of individual genes are obtained based on the training data. Next, we drop 195 genes whose sample variances are greater than $10^2$ or less than $10^{-2}$ after rescaling by a factor of $10^4$. Finally, to reduce the computational cost (which mainly comes from estimating the precision matrix), only top 200 genes (with the largest absolute values of the two sample $t$-statistics) are used for constructing different classification rules.  Figure \ref{fig2} illustrates the scores $\hat{T}^j=\frac{\exp(\hat{S}_j)}{1+\exp(\hat{S}_j)}$ estimated by LASS, Naive, Ebay and Lasso. For better illustration, the first 134 testing points are from ADCA group and the next 15 are from MPM group. 
	
	To achieve better separation, the values of the first 134 points (the next 15 points) should be high (low). We can see that LASS shows perfect separation of the two classes. While Ebay provides a clearer separation than Naive and Lasso, it is less effective than LASS. For this particular data set, if we choose to control the FSR at level $\alpha=0.1$,  then LASS makes no indecision. Thus, it makes sense to compare the misclassification rates under the classical setup. The results are presented in Table \ref{table4}, from which we can see that LASS has the best performance.

	\begin{table}
		\caption{\label{table4} {$\#$ of misclassifications} by each method}
		\centering
		\begin{tabular}{|l|l|l|l|l|l|l|}
			\hline
			& $\text{LASS}$  & $ \text{Naive}$   & $ \text{LPD}$ & $ \text{AdaLDA}$ & $ \text{Lasso}$ & $ \text{Ebay}$\\
			\hline
			$\text{Misclassification}$ & $\mathbf{0} ~ (\mathbf{0}\%)$    & $5 ~ (3.36\%)$          &  $1~(0.67\%) $   &  $1~(0.67\%) $    &  $5~(3.36\%) $  & $3~(2.01\%) $  \\
			\hline
			
		\end{tabular}
		
	\end{table}
	
	\subsection{p53 Mutants Data}\label{data:sec2}
	
	Finally we perform classification on p53 mutants data \citep{danziger2009predicting}, which consist of 16,772 tissue samples, and for each sample a $p=5,407$-dimensional vector is measured. Among the 16,772 samples, 143 are determined as ``active'' and the rest are determined as ``inactive''. We randomly select 100 active samples and 100 inactive samples as our training data, then take the rest 43 active samples and 50 random inactive samples as our testing set. To make the classification problem more difficult, an independent $\mathcal{N}(0,40)$ noise variable is added to each gene in both the training and testing sets. 
	
	We follow the previous pre-processing steps: (a) the training data are used to estimate the sample variances, (b) genes with variances greater than $10^2$ or smaller than $10^{-2}$ are dropped, and (c) only top 100 genes with the largest $t$-statistics are used. The experiment is repeated 50 times with results summarized in Tables \ref{table5} and \ref{table6}. 
	
	Table \ref{table5} contains the results under the conventional setup. LASS performs as well as the LPD and AdaLDA rules. But all methods have high misclassification rates. Hence we consider FSR control. We set the target FSR levels for both classes to be 0.1. In Table \ref{table6} we compare the FSR and power of different methods. We can see that LASS  effectively controls the FSR while Naive, Ebay and Lasso fail to do so. 
	
	\begin{table}
		\caption{\label{table5} Misclassification rates of different methods}
		\centering
		\begin{tabular}{|l|l|l|l|l|l|l|}
			\hline
			& $\text{LASS}$  & $ \text{Naive}$   & $ \text{LPD}$ & $ \text{AdaLDA}$ & $ \text{Lasso}$ & $ \text{Ebay}$\\
			\hline
			$\text{Misclassification}$ & $30.73\%$    & $40.24\%$          &  $32.34\% $   &  $\mathbf{30.28\%} $    &  $31.23\% $  & $31.57\% $  \\
			\hline
			
		\end{tabular}
		
	\end{table}

	\begin{table}
		\caption{\label{table6} FSR and power comparison.}
		\centering
		\begin{tabular}{|l|l|l|l|l|}
			\hline
			& $\text{LASS}$  & $ \text{Naive}$   & $ \text{Lasso}$ & $ \text{Ebay}$\\
			\hline
			$\text{FSR}^1$ & $\mathbf{10.38}\%$    & $41.90\%$          &  $17.81\%$  & $32.01\%$  \\
			\hline
			$\text{FSR}^2$ & $\mathbf{11.51}\%$    & $38.75\%$          &  $16.00\%$  & $30.88\%$  \\
			\hline
			$\text{Power}$ & $20.60\%$    & $59.38\%$          &  $19.25\%$  & $68.41\%$  \\
			\hline
			
		\end{tabular}
		
	\end{table}


	
	\bibliographystyle{chicago}
	\bibliography{myref}

\begin{thebibliography}{}

\bibitem[\protect\citeauthoryear{Angelopoulos, Bates, Cand{\`e}s, Jordan, and
  Lei}{Angelopoulos et~al.}{2021}]{Angetal21}
Angelopoulos, A.~N., S.~Bates, E.~J. Cand{\`e}s, M.~I. Jordan, and L.~Lei
  (2021).
\newblock Learn then test: Calibrating predictive algorithms to achieve risk
  control.
\newblock {\em arXiv:2110.01052\/}, Preprint.

\bibitem[\protect\citeauthoryear{Avella-Medina, Battey, Fan, and
  Li}{Avella-Medina et~al.}{2018}]{avella2018robust}
Avella-Medina, M., H.~S. Battey, J.~Fan, and Q.~Li (2018).
\newblock Robust estimation of high-dimensional covariance and precision
  matrices.
\newblock {\em Biometrika\/}~{\em 105\/}(2), 271--284.

\bibitem[\protect\citeauthoryear{Bates, Cand{\`e}s, Lei, Romano, and
  Sesia}{Bates et~al.}{2021}]{Batetal21}
Bates, S., E.~Cand{\`e}s, L.~Lei, Y.~Romano, and M.~Sesia (2021).
\newblock Testing for outliers with conformal p-values.
\newblock {\em arXiv:2104.08279\/}, Preprint.

\bibitem[\protect\citeauthoryear{Benjamini}{Benjamini}{2010}]{Ben10}
Benjamini, Y. (2010).
\newblock Simultaneous and selective inference: Current successes and future
  challenges.
\newblock {\em Biometrical Journal\/}~{\em 52\/}(6), 708--721.

\bibitem[\protect\citeauthoryear{Benjamini and Hochberg}{Benjamini and
  Hochberg}{1995}]{BenHoc95}
Benjamini, Y. and Y.~Hochberg (1995).
\newblock Controlling the false discovery rate: a practical and powerful
  approach to multiple testing.
\newblock {\em J. Roy. Statist. Soc. B\/}~{\em \textbf{57}}, 289--300.

\bibitem[\protect\citeauthoryear{Bickel and Levina}{Bickel and
  Levina}{2004}]{bickel2004some}
Bickel, P.~J. and E.~Levina (2004).
\newblock Some theory for fisher's linear discriminant function,naive bayes',
  and some alternatives when there are many more variables than observations.
\newblock {\em Bernoulli\/}~{\em 10\/}(6), 989--1010.

\bibitem[\protect\citeauthoryear{Bickel and Levina}{Bickel and
  Levina}{2008}]{bickel2008regularized}
Bickel, P.~J. and E.~Levina (2008).
\newblock Regularized estimation of large covariance matrices.
\newblock {\em The Annals of Statistics\/}~{\em 36\/}(1), 199--227.

\bibitem[\protect\citeauthoryear{Brown and Greenshtein}{Brown and
  Greenshtein}{2009}]{BroGre09}
Brown, L.~D. and E.~Greenshtein (2009).
\newblock Nonparametric empirical {B}ayes and compound decision approaches to
  estimation of a high-dimensional vector of normal means.
\newblock {\em The Annals of Statistics\/}~{\em 37}, 1685--1704.

\bibitem[\protect\citeauthoryear{Cai and Liu}{Cai and
  Liu}{2011}]{cai2011direct}
Cai, T. and W.~Liu (2011).
\newblock A direct estimation approach to sparse linear discriminant analysis.
\newblock {\em Journal of the American statistical association\/}~{\em
  106\/}(496), 1566--1577.

\bibitem[\protect\citeauthoryear{Cai, Liu, and Luo}{Cai
  et~al.}{2011}]{cai2011constrained}
Cai, T., W.~Liu, and X.~Luo (2011).
\newblock A constrained $l1$ minimization approach to sparse precision matrix
  estimation.
\newblock {\em Journal of the American Statistical Association\/}~{\em
  106\/}(494), 594--607.

\bibitem[\protect\citeauthoryear{Cai and Zhang}{Cai and
  Zhang}{2019}]{tony2019high}
Cai, T. and L.~Zhang (2019).
\newblock High dimensional linear discriminant analysis: optimality, adaptive
  algorithm and missing data.
\newblock {\em Journal of the Royal Statistical Society: Series B (Statistical
  Methodology)\/}~{\em 81\/}(4), 675--705.

\bibitem[\protect\citeauthoryear{Cai}{Cai}{2002}]{cai2002block}
Cai, T.~T. (2002).
\newblock On block thresholding in wavelet regression: Adaptivity, block size,
  and threshold level.
\newblock {\em Statistica Sinica\/}, 1241--1273.

\bibitem[\protect\citeauthoryear{Cai, Liu, Zhou, et~al.}{Cai
  et~al.}{2016}]{cai2016estimating}
Cai, T.~T., W.~Liu, H.~H. Zhou, et~al. (2016).
\newblock Estimating sparse precision matrix: Optimal rates of convergence and
  adaptive estimation.
\newblock {\em The Annals of Statistics\/}~{\em 44\/}(2), 455--488.

\bibitem[\protect\citeauthoryear{Danziger, Baronio, Ho, Hall, Salmon, Hatfield,
  Kaiser, and Lathrop}{Danziger et~al.}{2009}]{danziger2009predicting}
Danziger, S.~A., R.~Baronio, L.~Ho, L.~Hall, K.~Salmon, G.~W. Hatfield,
  P.~Kaiser, and R.~H. Lathrop (2009).
\newblock Predicting positive p53 cancer rescue regions using most informative
  positive (mip) active learning.
\newblock {\em PLoS computational biology\/}~{\em 5\/}(9), e1000498.

\bibitem[\protect\citeauthoryear{Dicker and Zhao}{Dicker and
  Zhao}{2016}]{dicker2016high}
Dicker, L.~H. and S.~D. Zhao (2016).
\newblock High-dimensional classification via nonparametric empirical bayes and
  maximum likelihood inference.
\newblock {\em Biometrika\/}~{\em 103\/}(1), 21--34.

\bibitem[\protect\citeauthoryear{Efron}{Efron}{2009}]{efron2009empirical}
Efron, B. (2009).
\newblock Empirical bayes estimates for large-scale prediction problems.
\newblock {\em Journal of the American Statistical Association\/}~{\em
  104\/}(487), 1015--1028.

\bibitem[\protect\citeauthoryear{Efron}{Efron}{2011}]{Efr11}
Efron, B. (2011).
\newblock Tweedie's formula and selection bias.
\newblock {\em Journal of the American Statistical Association\/}~{\em
  106\/}(496), 1602--1614.

\bibitem[\protect\citeauthoryear{Foygel and Drton}{Foygel and
  Drton}{2010}]{foygel2010extended}
Foygel, R. and M.~Drton (2010).
\newblock Extended bayesian information criteria for gaussian graphical models.
\newblock {\em Advances in neural information processing systems\/}~{\em 23}.

\bibitem[\protect\citeauthoryear{Friedman}{Friedman}{1989}]{Fri89}
Friedman, J.~H. (1989).
\newblock Regularized discriminant analysis.
\newblock {\em Journal of the American statistical association\/}~{\em
  84\/}(405), 165--175.

\bibitem[\protect\citeauthoryear{Gordon, Jensen, Hsiao, Gullans, Blumenstock,
  Ramaswamy, Richards, Sugarbaker, and Bueno}{Gordon
  et~al.}{2002}]{gordon2002translation}
Gordon, G.~J., R.~V. Jensen, L.-L. Hsiao, S.~R. Gullans, J.~E. Blumenstock,
  S.~Ramaswamy, W.~G. Richards, D.~J. Sugarbaker, and R.~Bueno (2002).
\newblock Translation of microarray data into clinically relevant cancer
  diagnostic tests using gene expression ratios in lung cancer and
  mesothelioma.
\newblock {\em Cancer research\/}~{\em 62\/}(17), 4963--4967.

\bibitem[\protect\citeauthoryear{Greenshtein and Park}{Greenshtein and
  Park}{2009}]{greenshtein2009application}
Greenshtein, E. and J.~Park (2009).
\newblock Application of non parametric empirical bayes estimation to high
  dimensional classification.
\newblock {\em Journal of Machine Learning Research\/}~{\em 10\/}(7).

\bibitem[\protect\citeauthoryear{Herbei and Wegkamp}{Herbei and
  Wegkamp}{2006}]{herbei2006classification}
Herbei, R. and M.~H. Wegkamp (2006).
\newblock Classification with reject option.
\newblock {\em The Canadian Journal of Statistics/La Revue Canadienne de
  Statistique\/}, 709--721.

\bibitem[\protect\citeauthoryear{James and Stein}{James and
  Stein}{1992}]{james1992estimation}
James, W. and C.~Stein (1992).
\newblock Estimation with quadratic loss.
\newblock In {\em Breakthroughs in statistics}, pp.\  443--460. Springer.

\bibitem[\protect\citeauthoryear{Koenker and Mizera}{Koenker and
  Mizera}{2014}]{KoeMiz14}
Koenker, R. and I.~Mizera (2014).
\newblock Convex optimization, shape constraints, compound decisions, and
  empirical {B}ayes rules.
\newblock {\em Journal of the American Statistical Association\/}~{\em
  109\/}(506), 674--685.

\bibitem[\protect\citeauthoryear{Lei}{Lei}{2014}]{lei2014classification}
Lei, J. (2014).
\newblock Classification with confidence.
\newblock {\em Biometrika\/}~{\em 101\/}(4), 755--769.

\bibitem[\protect\citeauthoryear{Liu and Luo}{Liu and Luo}{2015}]{liu2015fast}
Liu, W. and X.~Luo (2015).
\newblock Fast and adaptive sparse precision matrix estimation in high
  dimensions.
\newblock {\em Journal of multivariate analysis\/}~{\em 135}, 153--162.

\bibitem[\protect\citeauthoryear{Loh and Tan}{Loh and Tan}{2018}]{loh2018high}
Loh, P.-L. and X.~L. Tan (2018).
\newblock High-dimensional robust precision matrix estimation: Cellwise
  corruption under $\epsilon$-contamination.
\newblock {\em Electronic Journal of Statistics\/}~{\em 12\/}(1), 1429--1467.

\bibitem[\protect\citeauthoryear{Mai, Zou, and Yuan}{Mai
  et~al.}{2012}]{Maietal12}
Mai, Q., H.~Zou, and M.~Yuan (2012).
\newblock A direct approach to sparse discriminant analysis in ultra-high
  dimensions.
\newblock {\em Biometrika\/}~{\em 99\/}(1), 29--42.

\bibitem[\protect\citeauthoryear{Rava, Sun, James, and Tong}{Rava
  et~al.}{2021}]{Ravetal21}
Rava, B., W.~Sun, G.~M. James, and X.~Tong (2021).
\newblock A burden shared is a burden halved: A fairness-adjusted approach to
  classification.
\newblock {\em arXiv:2110.05720\/}, Preprint.

\bibitem[\protect\citeauthoryear{Rigollet and Tong}{Rigollet and
  Tong}{2011}]{RigTon11}
Rigollet, P. and X.~Tong (2011).
\newblock Neyman-pearson classification, convexity and stochastic constraints.
\newblock {\em Journal of Machine Learning Research\/}, 2831--2855.

\bibitem[\protect\citeauthoryear{Scott and Nowak}{Scott and
  Nowak}{2005}]{ScoNow05}
Scott, C. and R.~Nowak (2005).
\newblock A neyman-pearson approach to statistical learning.
\newblock {\em IEEE Transactions on Information Theory\/}~{\em 51\/}(11),
  3806--3819.

\bibitem[\protect\citeauthoryear{Shao, Wang, Deng, Wang, et~al.}{Shao
  et~al.}{2011}]{shao2011sparse}
Shao, J., Y.~Wang, X.~Deng, S.~Wang, et~al. (2011).
\newblock Sparse linear discriminant analysis by thresholding for high
  dimensional data.
\newblock {\em The Annals of statistics\/}~{\em 39\/}(2), 1241--1265.

\bibitem[\protect\citeauthoryear{Sun and Zhang}{Sun and
  Zhang}{2013}]{sun2013sparse}
Sun, T. and C.-H. Zhang (2013).
\newblock Sparse matrix inversion with scaled lasso.
\newblock {\em The Journal of Machine Learning Research\/}~{\em 14\/}(1),
  3385--3418.

\bibitem[\protect\citeauthoryear{Sun and Cai}{Sun and Cai}{2007}]{SunCai07}
Sun, W. and T.~T. Cai (2007).
\newblock Oracle and adaptive compound decision rules for false discovery rate
  control.
\newblock {\em J. Amer. Statist. Assoc.\/}~{\em \textbf{102}}, 901--912.

\bibitem[\protect\citeauthoryear{Sun and Wei}{Sun and Wei}{2011}]{SunWei11}
Sun, W. and Z.~Wei (2011).
\newblock Large-scale multiple testing for pattern identification, with
  applications to time-course microarray experiments.
\newblock {\em J. Amer. Statist. Assoc.\/}~{\em \textbf{106}}, 73--88.

\bibitem[\protect\citeauthoryear{Tibshirani, Hastie, Narasimhan, and
  Chu}{Tibshirani et~al.}{2003}]{Tibetal03}
Tibshirani, R., T.~Hastie, B.~Narasimhan, and G.~Chu (2003).
\newblock Class prediction by nearest shrunken centroids, with applications to
  dna microarrays.
\newblock {\em Statistical Science\/}, 104--117.

\bibitem[\protect\citeauthoryear{Wang, Banerjee, Hsieh, Ravikumar, and
  Dhillon}{Wang et~al.}{2013}]{wang2013large}
Wang, H., A.~Banerjee, C.-J. Hsieh, P.~Ravikumar, and I.~S. Dhillon (2013).
\newblock Large scale distributed sparse precision estimation.
\newblock In {\em NIPS}, Volume~13, pp.\  584--592.

\bibitem[\protect\citeauthoryear{Witten and Tibshirani}{Witten and
  Tibshirani}{2009}]{WitTib09}
Witten, D.~M. and R.~Tibshirani (2009).
\newblock Covariance-regularized regression and classification for high
  dimensional problems.
\newblock {\em Journal of the Royal Statistical Society: Series B (Statistical
  Methodology)\/}~{\em 71\/}(3), 615--636.

\bibitem[\protect\citeauthoryear{Yuan}{Yuan}{2010}]{yuan2010high}
Yuan, M. (2010).
\newblock High dimensional inverse covariance matrix estimation via linear
  programming.
\newblock {\em The Journal of Machine Learning Research\/}~{\em 11},
  2261--2286.

\end{thebibliography}

	\newpage
	
	\setcounter{page}{1} 
	
	\appendix
	\begin{center}\LARGE
		Online Supplementary Material for ``A Locally Adaptive Shrinkage Approach to False Selection Rate Control in High-Dimensional Classification''
	\end{center}
	
	\medskip
	
	This supplement contains the proofs of main {theorems, propositions and corollaries} (Section A), proofs of technical lemmas (Section B), and an argument establishing the asymptotic equivalence of FSR and mFSR (Section C). 

	\section{Proofs of main theorems, propositions and corollaries}

	We swap the proofs of Theorems \ref{thm3} and \ref{thm4} as the latter is simpler. Other proofs are arranged according to the orders in the text. 
	
	\subsection{Proof of Theorem \ref{thm:optimal_or}}\label{pf:thm1a}
	We only need to show $\delta_{OR}^j=2\mathbb{I}({T}^j<t^2_{OR})$ controls $\text{mFSR}^2$ at level $\alpha_2$ and maximizes $\mathbb{E}\left\{  \sum_{i=1}^{m}\mathbb{I}(\theta_j=2,\delta_j=2)    \right\}$. The same argument can be used to show that $\delta_{OR}^j=\mathbb{I}(1-{T}^j<t^1_{OR})$ controls $\text{mFSR}^1$ at level $\alpha_1$ and maximizes $\mathbb{E}\left\{  \sum_{i=1}^{m}\mathbb{I}(\theta_j=1,\delta_j=1)    \right\}$. The theorem is proved when we combine these two statements.

	The proof is divided into two parts. In part (a), we establish two properties of the classification rule $\pmb{\delta}^2_{}(t)=\{2\mathbb I (T^{j} < t): 1 \leq j \leq m\}$. We show that it produces $\text{mFSR}^2$ $< t$ for all $t\in(0,1)$ and that its $\text{mFSR}^2$ is monotonic in $t$. In part (b) we show that when the threshold is $t^2_{OR}$, the classification rule has $\text{mFSR}^2=\alpha_2$ and maximizes $\mathbb{E}\left\{  \sum_{i=1}^{m}\mathbb{I}(\theta_j=2,\delta_j=2)    \right\}$ amongst all valid classification rule with $\text{mFSR}^2\leq \alpha_2$.
	
	\medskip
	
	\noindent\textbf{Part(a).}  Consider classification rule $\{2\cdot\mathbb I (T^{j} < t): 1 \leq j \leq m\}$. Let $Q^2_{ }(t)=\alpha_t$ be its the $\text{mFSR}^2$ level. We first show that $\alpha_t < t$. Since ${T}^j = P(\theta_j = 1|\pmb{W}_j)$, then
	$$
	\mathbb{E}\left\{\sum_{j=1}^{m}\mathbb{I}(\theta_j=1,\delta_j=2)      \right\}=\mathbb{E}_{\pmb W} \left[ \left\{\sum_{j=1}^{m} \nonumber \mathbb{E}_{\pmb{\theta|W}}\mathbb{I}(\theta_j=1,\delta_j=2)      \right\} \right]=
	\mathbb{E}_{\pmb W}\left\{\sum_{i=1}^{m}T^{j}\mathbb{I}(\delta_j=2)\right\},
	$$
	where $\mathbb{E}$ is the expectation over $(\pmb{W, \theta})$, $E_{W}$ is the expectation over $\pmb W$, and  $E_{\pmb{\theta|W}}$ is the expectation over $\pmb \theta$ holding $\pmb W$ fixed. 
	Recall $Q^2_{OR}(t)=\alpha_t$ is the $\text{FSR}^2$ level of the classification rule 
	$
	\{2\cdot\mathbb I (T^{j} < t): 1 \leq i \leq m\},
	$
	we have
	\begin{equation}\label{eq:rewritemFdr}
		\mathbb{E}_{\pmb W}\left\{\sum_{j=1}^{m}(T^{j} - \alpha_{t})\mathbb{I}(T^{j} < t)\right\} = 0.
	\end{equation}
	This implies that $\alpha_t < t$. To see this, if $\alpha_t\geq t$, then $(T^{j} - \alpha_{t})\mathbb{I}(T^{j} < t)<0$, which contradicts the right hand side. 
	
	Next, we show that $Q^2_{}(t)$ is nondecreasing in  $t$. That is, letting $Q^2_{}(t_j) = \alpha_{t_j}$, if $t_1 < t_2$, then $\alpha_{t_1} \leq \alpha_{t_2}$. We argue by contradiction.  Suppose that $t_1 < t_2$ but $\alpha_{t_1} > \alpha_{t_2}$. Then 
	\begin{align}\label{eq:talpha2}
		&(T^{j} - \alpha_{t_2})\mathbb{I}(T^{j} < t_2)\\ \nonumber
		=&(T^{j} - \alpha_{t_1})\mathbb{I}(T^{j} < t_1)  +  (\alpha_{t_1} - \alpha_{t_2})  \mathbb I(T^{j} < t_1)+ (T^{j} - \alpha_{t_2}) \mathbb{I}(t_1\leq {T}^j <t_2)\\ \nonumber
		\geq &(T^{j} - \alpha_{t_1})\mathbb{I}(T^{j} < t_1)  +  (\alpha_{t_1} - \alpha_{t_2})  \mathbb I(T^{j} < t_1)+ (T^{j} - \alpha_{t_1}) \mathbb{I}(t_1\leq {T}^j <t_2).
	\end{align}
	By (\ref{eq:rewritemFdr}) we have $\mathbb{E}\left\{  \sum_{j=1}^{m} (T^{j} - \alpha_{t_1})\mathbb{I}(T^{j} < t_1)  \right\}=0$, together with the fact that $\alpha_{t_1}<t_1$ we have $\mathbb{E}\left\{  \sum_{j=1}^{m} (T^{j} - \alpha_{t_2})\mathbb{I}(T^{j} < t_2)  \right\}>0$, contradicting (\ref{eq:rewritemFdr}).

	\medskip
	
	\noindent\textbf{Part(b).} Define $t^2_{OR} = \sup_{t}\{t \in (0,1): Q^2_{}(t) \leq \alpha\}$. 
	By part (a), $Q^2_{}(t)$ is non--decreasing in $t$. By continuity, $Q^2_{}(T) = \alpha_2$. Next, consider the oracle rule $\pmb{\delta}^2_{OR} =(\delta_{OR}^{2,1},\ldots,\delta_{OR}^{2,m}) =\{2\mathbb I (T^{j} < t^2_{OR}): \ 1 \leq j \leq m\}$ and an arbitrary rule $\pmb{\delta}_{*} = (\delta_{*}^{1}, \cdots, \delta_{*}^{m})$ such that $\text{mFSR}^2_{\pmb{\delta}_{*}} \leq  \alpha_2$. Using the result in part (a), we have
	\begin{equation}\label{pf1}
		\mathbb{E}\left\{\sum_{j=1}^{m}(T^{j} - \alpha)\mathbb{I}(\delta_{OR}^{2,j}=2)\right\} = 0 \quad 
		\text{and} \quad \mathbb{E}\left\{\sum_{j=1}^{m}(T^{j} - \alpha)\mathbb{I}(\delta_{*}^j=2)\right\} \leq 0.
	\end{equation}
	Taking the difference of the two equations in (\ref{pf1}), we have
	\begin{equation}\label{eq:or.power1}
		\mathbb{E}\left\{\sum_{j=1}^{m}(T^{j} - \alpha)\mathbb{I}(\delta_{OR}^{2,j}=2) -
		(T^{j} - \alpha)\mathbb{I}(\delta_{*}^j=2) \right\}\geq 0.
	\end{equation}
	Next consider the transformation $f(x) = (x-\alpha)/(1-x) $. Note that $f'(x) = (1-\alpha)/(1-x)^2 >0$, $f(x)$ is monotonically increasing, the order is preserved by this transformation: if  $T^{i} < t^2_{OR}$ then $f(T^{i}) < f(t^2_{OR})$. This means we can rewrite the oracle rule as 
	$$
	\delta_{OR}^{2,i} = 2\mathbb I\left[\left\{ (T^{i} - \alpha)/(1-T^{i})\right\} < \lambda^2_{OR}\right],
	$$
	where $\lambda^2_{OR} =(t^2_{OR} - \alpha)/(1-t^2_{OR})$. It will be useful to note that, from part (a), we have $\alpha_{t_2} < t^2_{OR} < 1$, which implies that $\lambda^2_{OR} > 0$. Note that 
	\begin{equation}\label{eq:or.power2}
		\mathbb{E}\left[\sum_{j=1}^{m}\left\{\mathbb{I}(\delta_{OR}^{2,j}=2)-\mathbb{I}(\delta_{*}^{j}=2)\right\}\left\{(T^{j}-\alpha) - \lambda^2_{OR}(1-T^{j})\right\}\right] \leq 0
	\end{equation}
	To see this,  consider that if  $\mathbb{I}(\delta_{OR}^{2,j}=2)-\mathbb{I}(\delta_{*}^{j}=2)\neq 0$, then either  (i) $\mathbb{I}(\delta_{OR}^{2,j}=2)-\mathbb{I}(\delta_{*}^{j}=2)> 0$ or (ii) $\mathbb{I}(\delta_{OR}^{2,j}=2)-\mathbb{I}(\delta_{*}^{j}=2)< 0$ holds. If (i) holds, then $\delta_{OR}^{2,j} = 2$ and it follows that $\left\{(T^{j} - \alpha)/(1-T^{j})\right\} < \lambda^2_{OR}$. If (ii) holds, then $\delta_{OR}^{j} \neq 2$ and  $\left\{(T^{j} - \alpha)/(1-T^{j})\right\} \geq \lambda^2_{OR}$. In both cases, we have 
	$$
	\left\{\mathbb{I}(\delta_{OR}^{2,j}=2)-\mathbb{I}(\delta_{*}^{j}=2)\right\}\left\{(T^{j}-\alpha) - \lambda^2_{OR}(1-T^{i})\right\} \leq 0.
	$$ 
	Summing over all $m$ terms and taking the expectation yields \eqref{eq:or.power2}.
	
	Combining \eqref{eq:or.power1} and \eqref{eq:or.power2}, we obtain
	$$
	0 \leq \lambda^2_{OR} \mathbb{E} \left[\sum_{j=1}^{m}\left\{\mathbb{I}(\delta_{OR}^{2,j}=2) -\mathbb{I}( \delta_{*}^{j}=2)\right\}(1-T^{j} ) \right]
	$$
	Finally, since $\lambda^2_{OR} > 0$, it follows that
	$$ \mathbb{E} \left[\sum_{i=1}^{m}\left\{\mathbb{I}(\delta_{OR}^{j}=2) -\mathbb{I}( \delta_{*}^{j}=2)\right\}(1-T^{j} ) \right]\geq 0.$$
	\qed 
	
	\subsection{Proof of Theorem \ref{thm:optimal_or2}}
	As in the proof of Theorem \ref{thm:optimal_or}, we will show $\pmb{\delta}_{OR}^{*2}$=\{$\delta_{OR}^{*2,j}=2\mathbb{I}({T}^j<\hat{t}^2_{OR}),\ j=1,\ldots,m\}$ controls $\text{mFSR}^2$ and $\text{FSR}^2$ at level $\alpha_2$. Let $\pmb{\theta}=(\theta_1,\ldots, \theta_m)$. Note that
	$$
		\text{FSR}^2= \mathbb{E}\left[   \dfrac{ \sum_{j=1}^{m}\delta_{OR}^{*2,j}\mathbb{I}(\theta_j\neq 2 ) }{ \{ \sum_{j=1}^{m}\delta_{OR}^{*2,j}   \}\vee 1    }    \right]=\mathbb{E}_{\pmb{\delta}_{OR}^{*2}}\mathbb{E}_{\pmb{\theta}| \pmb{\delta}_{OR}^{*2} }\left[   \dfrac{ \sum_{j=1}^{m}\delta_{OR}^{*2,j}\mathbb{I}(\theta_j\neq 2 ) }{ \{ \sum_{j=1}^{m}\delta_{OR}^{*2,j}   \}\vee 1    }  \bigg| \pmb{\delta}_{OR}^{*2}  \right].
		$$
		
		By definition of the rule $\pmb{\delta}_{OR}^{*2}$, if $\sum_{j=1}^{m}\delta_{OR}^{*2,j}=k\geq 1 $ then $$\mathbb{E}_{\pmb{\theta}|\pmb{\delta}_{OR}^{*2}}\sum_{j=1}^{m}\delta_{OR}^{*2,j}\mathbb{I}(\theta_j\neq 2)=\sum_{i=1}^{m} \mathbb{P}(\theta_j\neq 2)\delta^{*2,j}_{OR}\leq k\alpha_2.$$
		It follows that 
		$
		\mathbb{E}_{\pmb{\theta}| \pmb{\delta}_{OR}^{*2} }\left[   \dfrac{ \sum_{j=1}^{m}\delta_{OR}^{*2,j}\mathbb{I}(\theta_j\neq 2 ) }{ \{ \sum_{j=1}^{m}\delta_{OR}^{*2,j}   \}\vee 1    }  \bigg| \pmb{\delta}_{OR}^{*2}  \right]\leq \alpha,
		$
		and $	\text{FSR}^2\leq \alpha_2$.
		
		Let $\mathcal{W}=(\pmb{W}_1,\ldots,\pmb{W}_m)$. For $\text{mFSR}^2$ we have
	\begin{align*}
		\frac{\mathbb{E}\left(\sum_{j=1}^m \mathbb{I}(\theta_j\neq2)\delta^{*2,j}_{OR}\right)}{\mathbb{E}\left(\sum_{j=1}^m \delta^{*2,j}_{OR}\right)}&=	\frac{\mathbb{E}_{\mathcal{W}}\mathbb{E}_{\pmb{\theta}|\mathcal{W}}\left\{\sum_{j=1}^m \mathbb{I}(\theta_j\neq2)\delta^{*2,j}_{OR}\right\}}{\mathbb{E}\left(\sum_{j=1}^m \delta^{*2,j}_{OR}\right)}\\
		&=\frac{\mathbb{E}_{\mathcal{W}}\mathbb{E}_{\pmb{\theta}|\mathcal{W}}\left\{(\sum_{j=1}^{m} \delta^{*2,j}_{OR}) \left(\sum_{j=1}^m \mathbb{I}(\theta_j\neq2)\delta^{*2,j}_{OR}/\sum_{j=1}^m \delta^{*2,j}_{OR}\right)\right\}}{\mathbb{E}\left(\sum_{j=1}^m \delta^{*2,j}_{OR}\right)}\\
		&=	\frac{\mathbb{E}_{\mathcal{W}}\left(\sum_{j=1} \delta^{*2,j}_{OR}\right) \mathbb{E}_{\mathcal{W}} \left\{\mathbb{E}_{\pmb{\theta}|\mathcal{W}}\left( \sum_{j=1}^m \mathbb{I}(\theta_j\neq2)\delta^{*2,j}_{OR}/\sum_{j=1} \delta^{*2,j}_{OR}\right)\right\}}{\mathbb{E}_{\mathcal{W}}\left(\sum_{j=1}^m \delta^{*2,j}_{OR}\right)}\\
		&= \mathbb{E}_{\mathcal{W}}\left\{\mathbb{E}_{\pmb{\theta}|\mathcal{W}}\left( \sum_{j=1}^m \mathbb{I}(\theta_j\neq2)\delta^{*2,j}_{OR}/\sum_{j=1} \delta^{*2,j}_{OR}\right)\right\}.
	\end{align*}
	The fourth equality holds because $\pmb{\delta}^{*2}_{OR}$ depends on $\pmb{\theta}$ only through $\mathcal{W}$. Now given $\mathcal{W}$ and that $\sum_{i} \delta^{*2,j}_{OR} = k$, we have
	\begin{eqnarray*}
	\mathbb{E}_{\mathcal{W}}\left\{E_{\pmb{\theta}|\mathcal{W}}\left( \sum_{j=1}^m \mathbb{I}(\theta_j\neq 2)\delta^{*2,j}_{OR}/\sum_{j=1}^m \delta^{*2,j}_{OR}\right)\right\} & = & \mathbb{E}_{\mathcal{W}}\left\{ \frac{\mathbb{E}_{\pmb{\theta}|\mathcal{W}}\left( \sum_{j=1}^m \mathbb{I}(\theta_j\neq 2)\right)\delta^{*2,j}_{OR}}{k} \right\}\\ & = & \mathbb{E}_{\mathcal{W}}\left( \frac{ \sum_{j=1}^m T^j\delta^{*2,j}_{OR}}{k} \right).
	\end{eqnarray*}
	By the definition of $\pmb{\delta}_{OR}^{*2}$, we have
	\begin{align*}
	 \mathbb{E}_{\mathcal{W}}\left( \frac{ \sum_{j=1}^m T^j\delta^{*2,j}_{OR}}{k} \right)\leq \alpha_2.
	\end{align*}
	Hence  $\pmb{\delta}_{OR}^{*2}$ satisfies $\text{mFSR}^2\leq \alpha_2$.	\qed

	\subsection{Proof of Proposition \ref{thm2}}
	Let
	$q_k(x):=\tilde{g}_{1k}(|x|)/\{g_{0}(|x|)+\tilde{g}_{1k}(|x|)\} $,
	where $g_{0}$ and $\tilde{g}_{1k}$ are the density function of $ \mathcal{N}\left(0,\frac{n_1+n_2}{n_1n_2}\right)$ and 
	$\mathcal{N}\left({\left\{(2+b)\sqrt{\sigma_{kk}}+\sqrt{(2+b)^2\sigma_{kk}+4}\right\}}\sqrt{\ninv\log p},\nn\right)$ respectively. We will assume without loss of generality that $d_k>0$.
	
	{We shall first prove the result when  the true variance $\sigma_{kk}$ is known and then argue that with probability greater than $1-pe^{-O(n)}$ the result still holds when $\sigma_{kk}$ is replaced by $\hat{\sigma}_{kk}$.}
	Let $g_{2k}$ be the density function of $\mathcal{N}\left(d_k,\sigma_{kk}\nn\right)$. Assume $d_k$ is strong, we investigate the asymptotic behavior of 
	\begin{eqnarray}\label{eq3}
		1-\mathbb{E}\{q_k(X)|X\sim g_{2k} \} &= &\int_{-\infty}^{\infty}\dfrac{g_{0}(|x|)g_{2k}(x)}{g_{0}(|x|)+ \tilde{g}_{1k}(|x|)}  dx \\\nonumber
		&=& \int_{-\infty}^{d_k-\frac{\epsilon}{4}\sqrt{\ninv\log p}}\dfrac{g_{0}(|x|)g_{2k}(x)}{g_{0}(|x|) +\tilde{g}_{1k}(|x|)}  dx \\ \nonumber && +\int_{d_k-\frac{\epsilon}{4}\sqrt{\ninv\log p}}^{\infty}\dfrac{g_{0}(|x|)g_{2k}(x)}{g_{0}(|x|)+ \tilde{g}_{1k}(|x|)}  dx.
	\end{eqnarray}
	Consider the first term, it is easy to see that 
	\begin{align*}
		\int_{-\infty}^{d_k-\frac{\epsilon}{4}\sqrt{\ninv\log p}}\dfrac{g_{0}(|x|)g_{2k}(x)}{g_{0}(|x|)+ \tilde{g}_{1k}(|x|)}  dx&\leq \int_{-\infty}^{d_k-\frac{\epsilon}{4}\sqrt{\ninv\log p}} g_{2k}(x)dx=O(p^{-\epsilon^2/(64\sigma_{kk})}).
	\end{align*}
	Since $d_k\in \mathcal G_1$, we have $x>\left(a_k/2+3\epsilon/4\right)\sqrt{\ninv\log p}$ if $\left(d_k-(\epsilon/4)\sqrt{\ninv\log p},\infty\right)$; on this interval we have   
	\begin{align*}
		\dfrac{\tilde{g}_{1k}(|x|)}{g_{0}(|x|)}&=\exp\left(\dfrac{2a_k|x|\sqrt{\ninv\log p}-a_k^2\ninv\log p}{2\nn}   \right)\\
		&\geq \exp\left(  \frac{1}{2}a_k(a_k/2+3\epsilon/4)\log p-\frac{1}{4}a_k^2\log p          \right)\\
		&\geq \exp\left(\dfrac{3a_k\epsilon}{8} \log p \right)\\
		&\geq p^{3a_k\epsilon/8}.
	\end{align*}
	It follows that 
	\begin{eqnarray*}
		\int_{(d_k-\epsilon/4)\sqrt{\ninv\log p}}^{\infty}\dfrac{g_{0}(|x|)g_{2k}(x)}{g_{0}(|x|)+ \tilde{g}_{1k}(|x|)}  dx & \leq & \sup_{x>(d_k-\epsilon/4)\sqrt{\ninv\log p}}\dfrac{1}{1+\tilde{g}_{1k}(x)/g_{0}(x)} \\ & \leq &  p^{-\textcolor{black}{3a_k}\epsilon/8}.
	\end{eqnarray*}
	Using \eqref{eq3}, we have $\mathbb{E}(q_k|\theta_k=1)=O(p^{-\epsilon_1})$, where $\epsilon_1>0$ is some constant.

	Next, when $d_k\in \mathcal G_3$,  we can similarly show that 
	$$\mathbb{E}\{q_k(X)|X\sim g_{2k}\}=\left|1- \int_{-\infty}^{\infty}\dfrac{g_{0}(|x|)g_{2k}(x)}{g_{0}(|x|)+ \tilde{g}_{1k}(|x|)}  dx  \right| =O(p^{-\epsilon_2}). $$

	Finally, consider the the case when $d_k\in\mathcal G_2$. We have
	\begin{align*}
		\int_{-\infty}^{\infty}\dfrac{g_{0}(|x|)g_{2k}(x)}{g_{0}(|x|)+ \tilde{g}_{1k}(|x|)}  dx  &=\int_{-\infty}^{\infty}  \dfrac{g_{2k}(x)}{1+\exp\left(\dfrac{2{a_k}\sqrt{\ninv\log p}|x|-{a_k}^2\ninv\log p}{2\nn}\right)}   dx\\
		&= \int_{-\infty}^{\infty}  \dfrac{g_{2k}(x)}{1+\exp\left(\frac{1}{\sqrt{2}}\textcolor{black}{a_k}|x|\sqrt{\frac{n_1n_2}{n_1+n_2}\log p}-\frac{{a_k}^2}{4}\log p\right)}   dx. 
	\end{align*}
	Let $\mathcal{A}_t=\left\{x: -t/\sqrt{\nt}+d_k\leq x \leq t/\sqrt{\nt}+d_k
	\right\}$,
	\begin{align*}
		&\int_{-\infty}^{\infty}  \dfrac{g_{2k}(x)}{1+\exp\left(\frac{1}{\sqrt{2}}\textcolor{black}{a_k}|x|\sqrt{\frac{n_1n_2}{n_1+n_2}\log p}-\frac{{a_k}^2}{4}\log p\right)} - g_{2k}(x) dx\\
		=&\int_{x\in \mathcal{A}_t^c} \dfrac{g_{2k}(x)}{1+\exp\left(\frac{1}{\sqrt{2}}\textcolor{black}{a_k}|x|\sqrt{\frac{n_1n_2}{n_1+n_2}\log p}-\frac{{a_k}^2}{4}\log p\right)} - g_{2k}(x)   dx\\
		+&\int_{x\in \mathcal{A}_t} \dfrac{g_{2k}(x)}{1+\exp\left(\frac{1}{\sqrt{2}}\textcolor{black}{a_k}|x|\sqrt{\frac{n_1n_2}{n_1+n_2}\log p}-\frac{{a_k}^2}{4}\log p\right)} - g_{2k}(x)   dx.
	\end{align*}
	Some algebra shows that 
	\begin{eqnarray}\label{eq:1}
		&\left| \int_{x\in \mathcal{A}_t^c} \dfrac{g_{2k}(x)}{1+\exp\left(\frac{1}{\sqrt{2}}\textcolor{black}{a_k}|x|\sqrt{\frac{n_1n_2}{n_1+n_2}\log p}-\frac{{a_k}^2}{4}\log p\right)}  - g_{2k}(x)   dx   \right| \nonumber\\
		\leq & \int_{x\in \mathcal{A}_t^c} g_{2k}(x)dx=O\left( e^{-\frac{t^2}{4\sigma_{kk}}}\right), \mbox{ and }
		\\			\nonumber		&\left| \int_{x\in \mathcal{A}_t} \dfrac{g_{2k}(x)}{1+\exp\left(\frac{1}{\sqrt{2}}\textcolor{black}{a_k}|x|\sqrt{\frac{n_1n_2}{n_1+n_2}\log p}-\frac{{a_k}^2}{4}\log p\right)}    dx\right|\\ \label{eq:2} 
		\leq& \left| \int_{x\in \mathcal{A}_t} \dfrac{\phi_{2k}(x)}{1+\exp\left(\frac{1}{2}\textcolor{black}{a_k}t\sqrt{\log p}+o(\log p)-\frac{\textcolor{black}{a_k}^2}{4}\log p\right)}- g_{2k}(x)   dx\right|.
	\end{eqnarray}
	\textcolor{black}{Take $t=\left(2+b/2\right)\sqrt{\sigma_{kk}\log p}$, we can see that \eqref{eq:1} is bounded by $o(1/p^{1+b})$. For (\ref{eq:2}) to be bounded by $o(1/p^{1+a})$ for some constant $a>0$, we need ${a_k}^2/4-(1+b/4){a_k}\sqrt{\sigma_{kk}}>1+c$ for some constant $c>0$. Some computation shows that $a_k> (2+b/2)\sqrt{\sigma_{kk}}+\sqrt{(2+b/2)^2\sigma_{kk}+4}$ can satisfy both requirements. Since we take
		$a_k= (2+b)\sqrt{\sigma_{kk}}+\sqrt{(2+b)^2\sigma_{kk}+4}$ we have \eqref{eq:2} is bounded by:}
	$$\left| \int_{x\in \mathcal{A}_t} \dfrac{g_{2k}(x)}{1+o(1/p^{1+\epsilon_3})}- g_{2k}(x)   dx   \right|= O(1/p^{1+\epsilon_3})\int g_{2k}(x)dx=O(1/p^{1+\epsilon_3}), $$
	where $\epsilon_3>0$ is some constant. Use the fact that $\int \phi_{2k} dx=1$ we conclude that 
	$$\mathbb{E}\{q_k(X)|X\sim g_{2k} \}=\left|1- \int_{-\infty}^{\infty}\dfrac{g_{0}(|x|)g_{2k}(x)}{g_{0}(|x|)+ \tilde{g}_{1k}(|x|)}  dx  \right| =O(1/p^{1+\epsilon_3}).$$
	Take $\gamma=\min(\epsilon_1,\epsilon_2,\epsilon_3)$, the results follow.
	
	Define $\hat{a}_k=(2+b)\sqrt{\hat{\sigma}_{kk}}+\sqrt{(2+b)^2\hat{\sigma}_{kk}+4}$.
	\textcolor{black}{From the above proof, we can see that if we would like the results to hold when we replace $a_k$ by $\hat{a}_k$, we need $x>\left(\hat{a}_k/2+3\epsilon'/4\right)\sqrt{\ninv\log p}$ on the interval $(d_k-(\epsilon/4)\sqrt{\ninv\log p},\infty)$ when $d_k$ is strong and $\hat{a}_k>(2+b/2)\sqrt{\sigma_{kk}}+\sqrt{(2+b/2)^2\sigma_{kk}+4}$ when $d_i$ is weak. Here $\epsilon'$ is some positive constant less than $\epsilon$ and depends solely on $\epsilon$. As $a_k$ satisfies these two conditions, we can choose a constant $d$ which depends solely on $\epsilon'$ and $b$ such that $\hat{a}_k$ satisfies these two conditions when $|\hat{a}_k-a_k|<d$.\\
		When the true variance is unknown, we use the pooled sample variance 
		$$\hat{\sigma}_{kk}=\frac{n_1-1}{n_1+n_2-2}\sum_{i=1}^{n_1}(X_{ik}-\bar{X}_k)^2+\frac{n_2-1}{n_1+n_2-2}\sum_{i=1}^{n_2}(Y_{ik}-\bar{Y}_k)^2
		$$
		to estimate $\sigma_{kk}$, and use $\hat{a}_k$ to estimate $a_k$. Since $\hat{a}_k$ is a continuous function of $\hat{\sigma}_{kk}$ and $\hat{a}_k=a_k$ when $\hat{\sigma}_{kk}=\sigma_{kk}$, we can choose a constant
		$\lambda$ small enough such that $|\hat{a}_k-a_k|<d$ when $|\hat{\sigma}_{kk}-\sigma_{kk}|<\epsilon_0^{-1}\lambda$. (Recall that $\epsilon_0^{-1}$ is the upper bound of $\sigma_{kk}$.) The constant $\lambda$ here depends solely on $d$ and $\epsilon_0$, which are fixed and independent of $n_1,\ n_2$ and $p$. Recall that since $X_{ik}$ are $i.i.d$ for $1\le i\le n_1$ and $Y_{ik}$ are $i.i.d$ for $1\le i\le n_2$ , we have $(n_1+n_2-2)\hat{\sigma}_{kk}/{\sigma_{kk}}\sim \chi^2_{n_1+n_2-2}$. Lemma 1 in \cite{foygel2010extended} and Lemma 4 in \cite{cai2002block} proved the following concentration inequality for $\chi^2$ random variable. For $n\ge 4\lambda^{-2}+1$, we have
		\begin{eqnarray*}
			P\left\{ \chi^2_n >n(1+\lambda)\right\} & \le & \frac{1}{\lambda\sqrt{\pi n}}e^{-\frac{n}{2}(\lambda-\log(1+\lambda))};\\
			P\left\{ \chi^2_n <n(1-\lambda)\right\} & \le &  \frac{1}{\lambda\sqrt{\pi (n-1)}}e^{-\frac{n-1}{2}(\lambda+\log(1-\lambda))}.
		\end{eqnarray*}
		Noting that $\lambda$ is a constant independent of $n_1,\ n_2$ and $p$, we apply the above results to conclude that 
		\begin{align*}
			& P\left(\left|\frac{\hat{\sigma}_{kk}}{\sigma_{kk}}-1\right|<\lambda\right) \\
			\ge& 1-\frac{1}{\lambda\sqrt{\pi (n_1+n_2-2)}}e^{-\frac{n_1+n_2-2}{2}(\lambda-\log(1+\lambda))}-\frac{1}{\lambda\sqrt{\pi (n_1+n_2-3)}}e^{-\frac{n_1+n_2-3}{2}(\lambda+\log(1-\lambda))}\\
			=& 1-e^{-\frac{n_1+n_2-2}{2}(\lambda-\log(1+\lambda))+O(\log{(n_1+n_2)})}-e^{-\frac{n_1+n_2-3}{2}(\lambda+\log(1-\lambda))+O(\log{(n_1+n_2)})}.\\
			=& 1-e^{-O(n_1+n_2)}
		\end{align*}
		Since $P\left(\bigcup\limits_{i=1}^{p}\left|\hat{\sigma}_{kk}-\sigma_{kk}\right|>\epsilon_0^{-1}\lambda\right)\le \sum_{i=1}^pP\left(\left|\hat{\sigma}_{kk}-\sigma_{kk}\right|>\epsilon_0^{-1}\lambda\right)
		\le \sum_{i=1}^pP\left(\left|\hat{\sigma}_{kk}-\sigma_{kk}\right|>\sigma_{kk}\lambda\right)$, we have with probability greater than $1-pe^{-O(n_1+n_2)}$, $\left|\hat{\sigma}_{kk}-\sigma_{kk}\right|<\epsilon_0^{-1}\lambda$ for all $1\le i\le p$.}\\
	\textcolor{black}{Combining with the fact that $\mathbb{E}(q_k)$ is bounded, the theorem follows.}\qed
	
	\subsection{Proof of Theorem \ref{thm3}}
	\setcounter{equation}{0}
	Define $\text{ECC}^2_{\pmb{\delta}}=\EE\left\{   \sum_{j=1}^{m}\mathbb{I}(\delta_j=2,\theta_j=2)    \right\}$.
	We will prove $\text{mFSR}^2_{\hat{\pmb{\delta}}}=\text{mFSR}^2_{\pmb{\delta}_{OR}}+o(1)$ and $\text{ECC}^2_{\hat{\pmb{\delta}}}/\text{ECC}^2_{ \pmb{\delta}_{OR }}=1+o(1)$. Then by the same argument one can show $\text{mFSR}^1_{\hat{\pmb{\delta}}}=\text{mFSR}^1_{\pmb{\delta}_{OR}}+o(1)$ and $\text{ECC}^1_{\hat{\pmb{\delta}}}/\text{ECC}^1_{ \pmb{\delta}_{OR }}=1+o(1)$, then the theorem follows.

	We begin with a summary of notation used throughout the proof:
	\begin{itemize}
		\item $Q(t) = m^{-1}\left\{\sum_{j=1}^m (\Tor^{ j}-\alpha_2) \mathbb{I}(\Tor^{ j} < t)\right\}$. 
		\item $\hQ(t) = m^{-1}\left\{\sum_{j=1}^m (\hat{T}^{ j}-\alpha_2) \mathbb{I}(\hat{T}^{ j} < t)\right\}$. 
		\item $Q_{\infty}(t) = \mathbb{E}\{(\Tor-\alpha_2)\mathbb{I}(\Tor<t)\}$. 
		\item $t_{\infty} = \sup\{t : Q_{\infty}(t) \leq 0\}$ is the ``ideal" threshold.
	\end{itemize}
	Note that  $\Tor^j$ is a function of $\pmb{W}_j$ only. Since each $\pmb{W}_j$ are iid, it follows that $\Tor^j$ are also iid. Without loss of generality, assume the first $s_1$ signals are strong, the next $s_2$ signals are moderate and the rest are weak. We break the proof into 2 cases:\\
	Case 1: $\sum_{k=1}^{s_1}d^2_k<\infty$ as $n,p\rightarrow \infty$.\\
	Case 2: $\sum_{k=1}^{s_1}d^2_k\rightarrow\infty$ as $n,p\rightarrow \infty$.
	
	\subsubsection{Proof of case 1}
	We first show that $\text{mFSR}^2_{\hat{\pmb{\delta}}}=\text{mFSR}^2_{\pmb{\delta}_{OR}}+o(1)$.  We define a continuous version of $\hQ(t)$ using the following procedure: If $t_1$ and $t_2$ are two adjacent points of discontinuity, on the interval $[t_1,t_2]$, 
	$$ \hQ_{C}(t)= \dfrac{t-t_2}{t_1-t_2}\hQ(t_1)+\dfrac{t-t_1}{t_2-t_1}\hQ(t_2).$$ 
	As $t_\infty$ is proved to be larger than $\alpha_1$ in section \ref{pf:thm1a}, it is easy to verify that $\hQ_{C}(t)$ is continuous and monotone on the interval $[\alpha_1,1)$. Hence, its inverse $\hQ_{C}^{ -1}$ is well--defined, continuous, and monotone on the interval $[\alpha,1)$.
	
	Next, we shall show the following two results in turn: (i)  $\hQ(t) \overset{p}\rightarrow Q_{\infty}(t)$ and (ii) $\hQ_{C}^{-1}(0) \overset{p}\rightarrow t_{\infty}$. We shall need the following two lemmas, which are proved later:
	\begin{lemma}\label{lemma1}
		$\mathbb{E}(\hat{T}_{}-T)^2\rightarrow 0$.
	\end{lemma}
	\begin{lemma}\label{lemma:2}
		Let $V_j = (\Tor^{ j}-\alpha_2) \mathbb{I}(\Tor^{ j} < t)$
		and 
		$\hat{V}_j = (\hat{T}^{ j}-\alpha_2)\mathbb{I}(\hat{T}^{ j} < t)$.
		Then $\mathbb{E}\left(\hat{V}_j - V_j \right)^2 = o(1)$.
	\end{lemma}
	
	To show (i), note that $Q(t) \overset{p}\rightarrow Q_{\infty}(t)$ by the WLLN, so that we only need to establish that  $\hQ(t) \overset{p}\rightarrow Q(t)$.
	Let $S_m = \sum_{j=1}^m \left(\hat{V}_j - V_j\right)$. 
	By Lemma \ref{lemma:2} and the Cauchy-Schwartz inequality, $\mathbb{E}\left\{\left(\hat{V}_i-V_i\right)\left(\hat{V}_j-V_j\right)\right\} = o(1)$.
	%
	%
	It follows that 
	\begin{align*}
		Var\left( m^{-1} S_m   \right) = &  m^{-2} Var(S_m) \leq m^{-2}\sum_{j=1}^{m} \mathbb{E}\left\{ \left( \hat{V}_j - V_j\right)^2 \right\} \\
		& +O\left(\frac{1}{m^2}\sum_{i,j:i\neq j} \mathbb{E}\left\{\left(\hat{V}_i-V_i\right)\left(\hat{V}_j-V_j\right)\right\}\right)\\
		=& o(1).
	\end{align*}
	By Lemma \ref{lemma:2}, $\mathbb{E}(m^{-1}S_m)\rightarrow 0$, applying Chebyshev's inequality, we obtain $m^{-1}S_m = \hQ^{}(t) - Q_{}(t) \overset{p}\rightarrow 0$. Hence (i) is proved.\\
	Next, we show (ii). Since  $\hQ_{C}(t)$ is continuous, for any $\varepsilon>0$, we can find $\eta>0$ such that 
	$\left|\hQ_{C}^{ -1}(0) - \hQ_{C}^{ -1}\left\{\hQ_{C}^{}\left(t_{\infty}^{}\right)\right\}\right| < \varepsilon$ 
	if 
	$\left|\hQ_{C}^{}\left(t_{\infty}^{}\right) \right|< \eta$. It follows that
	\[
	P\left\{ \left|\hQ_{C}^{}\left(t_{\infty}^{}\right) \right|> \eta\right\} \geq P\left\{\left|\hQ_{C}^{ -1}(0) - \hQ_{C}^{ -1}\left\{\hQ_{C}^{}\left(t_{\infty}^{}\right)\right\}\right| > \varepsilon\right\}.
	\]
	Lemma \ref{lemma1} and the WLLN imply that $\hQ_{C}^{}(t) \overset{p}\rightarrow Q_{\infty}^{}(t).$ Note that $Q_{\infty}^{}\left(t_{\infty}^{}\right) = 0$. Then,
	\[
	P\left(\left|\hQ_{C}^{}\left(t_{\infty}^{}\right) \right|>\eta\right) \rightarrow 0.
	\]
	Hence, we have
	\beq\label{eq:ddConvergenceToIdealThresh}
	\hQ_{C}^{ -1}(0) \overset{p}\rightarrow \hQ_{C}^{ -1} \left\{\hQ_{C}^{}\left(t_{\infty}^{}\right)\right\} = t_{\infty}^{},
	\eeq
	completing the proof of (ii). Notice that $Q_{\infty}^{}(t)$ is continuous by construction, by (i) we also have $\hQ(t)\overset{p}\rightarrow \hQ_{C}(t)$.\\
	We can similarly define the continuous version of $Q^{}(t)$ as $Q^{}_{C}(t)$ and the  corresponding threshold as $Q_{C}^{ -1}(0)$. Write $\hat{\pmb{\delta}}=\hat{\pmb{\delta}}^1+\hat{\pmb{\delta}}^2$, where $\hat{\pmb{\delta}}^1$ is of the form $\mathbb{I} \left\{1-\hat{T}^{ j} _{}\leq \beta_1\right\}$ and $\hat{\pmb{\delta}}^2$ is of the form $2\mathbb{I} \left\{\hat{T}^{ j} _{}\leq \beta_2\right\}$ for some $\beta_1,\beta_2>0$. Similarly, write $\pmb{\delta}_{}=\pmb{\delta}^1_{}+\pmb{\delta}^2_{}$, where $\pmb{\delta}^1_{}$ is of the form $\mathbb{I} \left\{1-T^{ j} _{}\leq t^1_{OR}\right\}$ and $\pmb{\delta}^2_{OR}$ is of the form $2\mathbb{I} \left\{T^{ j} _{}\leq t^2_{OR}\right\}$ for some $\beta_1,\beta_2>0$. 
	Then by construction, we have 
	$$
	\hat{\pmb{\delta}}^2 = \left[2\mathbb{I} \left\{\hat{T}^{ j} _{}\leq \hQ_{C}^{ -1}(0)\right\}: 1 \leq j \leq m\right]
	\quad 
	\text{and}
	\quad
	\pmb{\delta}_{OR}^{2}= \left[2\mathbb{I} \left\{\Tor^{j} \leq Q_{C}^{ -1}(0)\right\}: 1 \leq i \leq m\right].
	$$
	Also, following the previous arguments, we can show that
	\beq\label{eq:orConvergenceToIdealThresh}
	Q_{C}^{ -1}(0) \overset{p}\rightarrow  t_{\infty}^{}.
	\eeq
	According to \eqref{eq:ddConvergenceToIdealThresh} and \eqref{eq:orConvergenceToIdealThresh}, we have
	\beq\label{eq:ddthreshConvergenceToOr}
	\hQ_{C}^{ -1}(0) = Q_{C}^{ -1}(0) + o_{p}(1).
	\eeq
	Note that the $\text{mFSR}^2$ level of $\pmb{\delta}_{OR}^{}$ and $\hat{\pmb{\delta}}$ are
	\[
	\text{mFSR}^2_{\pmb{\delta}_{OR}}  = \frac{P_{\theta_j=1}\left\{\Tor^{ j} \leq Q_{C}^{ -1}(0) \right\}}{P\left\{\Tor^{ j} \leq Q_{C}^{ -1}(0) \right\}}
	\quad
	\text{and}
	\quad
	\text{mFSR}^2_{\hat{\pmb{\delta}}^{}} = \frac{P_{\theta_j=1}\left\{\hat{T}^{ j} \leq q_{C}^{ -1}(0) \right\}}{P\left\{\hat{T}^{ j} \leq q_{C}^{ -1}(0) \right\}}.
	\]
	
	From Lemma \ref{lemma1}, $\hat{T}^{ j} \overset{p}\rightarrow \Tor^{ j}$, \eqref{eq:orConvergenceToIdealThresh}, and by the continuous mapping theorem,
	$\text{mFSR}^2_{\hat{\pmb{\delta}}^{}}   = \text{mFSR}^2_{\pmb{\delta}_{OR}} + o(1)=\alpha_2+o(1)$. By the asymptotic equivalence between mFSR and FSR (Appendix \ref{fsrmsfr}), the desired result follows.\\
	Next we show that $\text{ECC}^2_{\hat{\pmb{\delta}}^{}} $ /$\text{ECC}^2_{\pmb{\delta}_{OR}}  =1 + o(1)$. By definition, 
	$$\text{ECC}^2_{\hat{\pmb{\delta}}^{}} =\mathbb{E}\left\{\sum_{j=1}^{m}\mathbb{I}(\hat{T}^{ j} \leq \hQ_{C}^{ -1}(0) )(1-\Tor^i) \right\}\quad
	\text{and}
	\quad
	\text{ECC}^2_{\pmb{\delta}_{OR}} =  \mathbb{E}\left\{\sum_{j=1}^{m}\mathbb{I}(\Tor^{ j} \leq Q_{C}^{ -1}(0) )(1-\Tor^j) \right\}.$$ 
	Using the fact that  $\hat{T}^{ j} \overset{p}\rightarrow \Tor^{ j}$ and $\hQ_{C}^{ -1}(0)   \overset{p}\rightarrow Q_{C}^{ -1}(0)$, the result follows.

	\subsubsection{Proof of case 2}
	From the proof of Lemma 1, we have:
	\begin{equation}\label{eq:case2difference}
		\mathbb{E}\left( S^\pi_{j}- \hat{S}_j  \right)^2=O(s_1\nn)+o(\sum_{k=1}^{s_1}d^2_k)+O(\sum_{k=1}^{s_1}d_k\nn)+o(1).
	\end{equation}
	Since $\sum_{k=1}^{s_1}d^2_k\geq cs_1\nn\log p$ for some $c>0$ and $\mathbb{E}( S^\pi_j) \propto \sum_{k=1}^{s_1}d_k^2$,
	by (\ref{eq:case2difference}) 
	$$\mathbb{E}\left( S^\pi_{j}- \hat{S}_j  \right)^2\bigg/\sum_{k=1}^{s_1}d_k^2=O\left(\dfrac{1}{\log p}\right).$$ 
	It follows that $\mathbb{E}\sum_{k=1}^{s_1}\{q_k(\bar{X}_k-\bar{Y}_k)\}^2\propto \mathbb{E}(|S^\pi_j|) \propto \sum_{k=1}^{s_1}d_k^2$.
	If $\sum_{k=1}^{s_k}d_i^2\rightarrow \infty$, then $\hat{T}^j\rightarrow 0$ or  $\hat{T}^j\rightarrow \infty$. In either case, we have perfect separation between the two classes asymptotically. The theorem follows. \qed

	\subsection{Proof of Theorem \ref{thm4}}
	Again, without loss of generality, assume the first $s_1$ signals are strong, the next $s_2$ signals are moderate and the rest are weak. Consider the same model as described in section 2, but replace $\pmb{\mu}_1$ and $\pmb{\mu}_2$ by
	$\tilde{\pmb{\mu}}_1=(\mu_{11},...,\mu_{1s_1},0,...,0)^t $ and
	$\tilde{\pmb{\mu}}_2=(\mu_{21},,...,\mu_{2s_1},0,...,0)^t $ respectively.
	Let $\tilde{\pmb{W}}_j=\pmb{W}_j-\pmb{\mu}_1+\tilde{\pmb{\mu}}_1 $ when $\theta_j=1$ and $\tilde{\pmb{W}}_j=\pmb{W}_j-\pmb{\mu}_2+\tilde{\pmb{\mu}}_2 $ when $\theta_j=2$. It is then clear that 
	$$
	\tilde{\pmb{W}}_j\sim \mathbb{I}(\theta_j=1)\mathcal{N}(\tilde{\pmb{\mu}}_1,\Sigma)+ \mathbb{I}(\theta_j=2)\mathcal{N}(\tilde{\pmb{\mu}}_2,\Sigma).
	$$
	Denote $\tilde{\pmb{d}}=\tilde{\pmb{\mu}}_1-\tilde{\pmb{\mu}}_2$, define 
	$$Z_j:=Z(\tilde{\pmb{W}}_j):=\bigg(\tilde{\pmb{W}}_{j}-\dfrac{ \tilde{\pmb{\mu}}_{1}+\tilde{\pmb{\mu}_{2}}}{2}  \bigg)^\top  \Sigma^{-1}\tilde{\pmb{d}},$$
	$Z^j_{OR}:=Z_{OR}(\tilde{\pmb{W}}_j)=\exp(Z_j)/(\exp(Z_j)+1).$
	Consider the decision rule:
	\begin{equation*}
		\delta_Z^j=2\mathbb{I}\{ Z^j_{OR}\leq \min(Z_{OR}^{(k_4)},0.5)\}+\mathbb{I}\{ 1-Z^j_{OR}\leq \min(1-Z_{OR}^{(m-k_3)},0.5)\},
	\end{equation*}\label{decision}
	where
	\begin{equation*}
		k_3= \inf\left\{j: \frac{1}{j+1} \sum_{i=0}^j (1-Z_{OR}^{(m-i)})\leq\alpha_1\right\} \quad
		\text{and}
		\quad
		k_4= \sup\left\{j: \frac 1 j \sum_{i=1}^j Z_{OR}^{(i)}\leq\alpha_2\right\}.
	\end{equation*}
	By the proof of theorem 1 and theorem 3, $\pmb{\delta}_Z=(\delta^1_Z,\ldots, \delta^m_Z)$ controls $\text{mFSR}^1$ and $\text{mFSR}^2$ at level $\alpha_1$ and $\alpha_2$ respectively.
	Now notice that
	\begin{align*}
		& \mathbb{E}\left(Z_j-\hat{S}_j\right)^2\\
		=&\mathbb{E}\left\{\left(\tilde{\pmb{W}}_j-\dfrac{\tilde{\pmb{\mu}}_1+\tilde{\pmb{\mu}}_2}{2}\right)^\top\Sigma^{-1}\tilde{\pmb{d}}-\left(\pmb{W}_j-\dfrac{\bar{\pmb X}+\bar{\pmb Y}}{2}\right)^\top\hat{\Sigma}^{-1}\hat{\pmb{d}}\right\}^2\\
		=&O\left[\mathbb{E}\left\{\left(\pmb{W}_i-\dfrac{\pmb{\mu}_1+\pmb{\mu}_2}{2}\right)^\top\Sigma^{-1}(\tilde{\pmb{d}}-\hat{\pmb{d}})\right\}^2\right.
		+\mathbb{E}\left\{\left(\tilde{\pmb{W}}_j-\pmb{W}_j+\dfrac{\bar{\pmb X}+\bar{\pmb Y}}{2}-\dfrac{\pmb{\mu}_1+\pmb{\mu}_2}{2}\right)^\top\Sigma^{-1}\hat{\pmb{d}}\right\}^2\\
		+&\left. \mathbb{E}\left\{\left(\pmb{W}_i-\dfrac{\bar{\pmb X}+\bar{\pmb Y}}{2}\right)^\top(\Sigma-\hat{\Sigma}^{-1})\hat{\pmb{d}}\right\}^2\right]\\
		=&\uppercase\expandafter{\romannumeral1}+\uppercase\expandafter{\romannumeral2}+\uppercase\expandafter{\romannumeral3}
	\end{align*}
	For $\uppercase\expandafter{\romannumeral2}$, note that
	\begin{align*}
		&\mathbb{E}\left\{   \left(\tilde{\pmb{W}}_j-\pmb{W}_j+ \frac{\tilde{\pmb{\mu}}_1+\tilde{\pmb{\mu}}_2}{2}-\frac{\bar{\pmb X}+\bar{\pmb Y}}{2}\right)^\top\Sigma^{-1}\hat{\pmb{d}}\right\}^2\\
		=&O\left[\mathbb{E}\left\{\hat{\pmb{d}}^\top(\Sigma^{-1})^t\frac{1}{4}\nn\Sigma\Sigma^{-1}\hat{\pmb{d}}\right\}+  \left\{\lambda_{max}(\Sigma^{-1})p^{-\gamma}\log p\right\}^2   \right]\\
		=&O\left[\mathbb{E}\left\{\hat{\pmb{d}}^\top(\Sigma^{-1})^\top\frac{1}{4}\nn\hat{\pmb{d}}\right\}+\left\{\lambda_{max}(\Sigma^{-1})p^{-\gamma}\log p\right\}^2 \right]\\
		=& O\left[\mathbb{E}\left(\nn\lambda_{max}(\Sigma^{-1})||\pmb{d}-\hat{\pmb{d}}||^2\right)+\mathbb{E}\left(\nn\lambda_{max}(\Sigma^{-1})||\pmb{d}||^2\right)+\left\{\lambda_{max}(\Sigma^{-1})p^{-\gamma}\log p\right\}^2 \right]\\
		=&O\left[\mathbb{E}\nn\lambda_{max}(\Sigma^{-1})(||\pmb{d}-\hat{\pmb{d}}||^2+||\pmb{d}||^2)+\left\{\lambda_{max}(\Sigma^{-1})p^{-\gamma}\log p\right\}^2 \right].
	\end{align*}  
	By (A1) $\left\{\lambda_{max}(\Sigma^{-1})p^{-\gamma}\log p\right\}^2 \rightarrow 0$, use the same computation in the proof of lemma \ref{lemma1}, we have 
	$$\mathbb{E}\left( Z_j-\hat{S}_j\right)^2=
	O\left( \mathbb{E}\sum_{k=1}^{p}\{q_k(\bar{X}_k-\bar{Y}_k)-\tilde{d}_k\}^2  \right)+o(1),$$ 	$$\mathbb{E}\sum_{k=1}^{s_1}\{q_k(\bar{X}_k-\bar{Y}_k)-\tilde{d}_k\}^2=O(s_1\nn)+o(\sum_{k=1}^{s_1}d^2_k)+O(\sum_{k=1}^{s_1}d_k\nn)+o(1) .$$
	We know that
	$$ \mathbb{E}\sum_{i=s_1+1}^{s_1+s_2} \left\{q_k(\bar{X}_k-\bar{Y}_k)\right\}^2=O(s_2)O(p^{-\gamma})O\left(\nn\log p\right) 
	=O(p^{-\gamma}\log p)\rightarrow 0,$$
	$$
	\mathbb{E}\sum_{k=s_2+1}^{p} \{q_k(\bar{X}_k-\bar{Y}_k)\}^2=O(p)O(p^{-1-\gamma})o\left(\nn\log p\right)\rightarrow 0. 
	$$
	Thus, 
	\begin{equation}\label{eq:theo4case2}
		\mathbb{E}\left( Z_j-\hat{S}_j\right)^2=O(s_1\nn)+o(\sum_{k=1}^{s_1}d^2_k)+O(\sum_{k=1}^{s_1}d_k\nn)+o(1).
	\end{equation}
	Since $\sum_{k=1}^{s_1}d^2_k\geq cs_1\log p \nn$ for some $c>0$ and $\mathbb{E}(Z_j) \propto \sum_{k=1}^{s_1}d_k^2$,
	by (\ref{eq:theo4case2}) 
	$$\mathbb{E}\left( Z_{j}- \hat{S}_j  \right)^2\bigg/\sum_{k=1}^{s_1}d_k^2=O\left(\dfrac{1}{\log p}\right)$$
	It follows that $\mathbb{E}\sum_{k=1}^{s_1}\{q_k(\bar{X}_k-\bar{Y}_k)\}^2\propto \mathbb{E}(|Z_j|) \propto \sum_{k=1}^{s_1}d_k^2$.
	If $\sum_{k=1}^{s_1}d_k^2\rightarrow \infty$, then $\hat{T}^j\rightarrow 0$ or  $\hat{T}^j\rightarrow 1$. In either case, we have perfect separation between the two classes asymptotically. The theorem follows trivially. If $\sum_{k=1}^{s_1}d_k^2<\infty$, then
	$$
	\left(\mathbb{E}\left|\hat{T}^j-Z^j_{OR}\right|\right)^2=e^{O(\sum_{k=1}^{s_1}d^2_k)}\left\{O\left(s_1\nn\right)+O(\sum_{k=1}^{s_1}d^2_kp^{-2\gamma}) +O(p^{-\gamma}\log p)\right\}.
	$$
	Since $\sum_{k=1}^{s_1}d^2_k\geq c s_1\log p\nn$ and 
	$\sum_{k=1}^{s_1}d^2_k<\infty$, it follows that $e^{O(\sum_{k=1}^{s_1}d^2_k)}<\infty$ \\
	and $O\left(s_1\nn\right)+O(\sum_{k=1}^{s_1}d^2_kp^{-2\gamma}) +O(p^{-\gamma}\log p)\rightarrow 0$. Thus, $\mathbb{E}\left(\hat{T}^j-Z^j_{OR}\right)^2\rightarrow 0.$ The theorem follows by using the same argument presented in the proof of theorem \ref{thm3}.
	\qed
	\subsection{Proof of Corollary \ref{coro1}}
	Note that $\min\{T^j,1-T^j\}\leq 0.5$. Hence, if we choose $\alpha_1=\alpha_2=0.5$, $\pmb{\delta}_{OR}$ makes no indecision. Since $\pmb{\delta}_{OR}$ has the highest ECC among all rules that controls $\text{mFSR}^1$ and $\text{mFSR}^2$ at level 0.5, and $\pmb{\delta}^F$ also controls $\text{mFSR}^1$ and $\text{mFSR}^2$ at level 0.5, it follows that $\text{ECC}_{\pmb{\delta}_{OR}} \geq\text{ECC}_{\pmb{\delta}^{F}} $. On the other hand, $\pmb{\delta}^{F}$ has the lowest risk among all rules that do not make indecision and risk equals to $1-\text{ECC}/m$. It follows that $\text{ECC}_{\pmb{\delta}_{OR}} \leq\text{ECC}_{\pmb{\delta}^{F}} $. Hence, $\text{ECC}_{\pmb{\delta}_{OR}} =\text{ECC}_{\pmb{\delta}^{F}} $. By Theorem \ref{thm3}, we have  $\text{ECC}_{\pmb{\delta}_{OR}} /\text{ECC}_{\hat{\pmb{\delta}}}\rightarrow 1 $, it follows that $R(\hat{\pmb{\delta}})\rightarrow R(\pmb{\delta}^F)$.\qed

	\section{Proof of Technical Lemmas}
	Appendix B contains proofs of techincal lemmas.
	We will assume without loss of generality, the first $s_1$ signals are strong, the next $s_2$ signals are moderate and the rest are weak.
	
	\subsection{Proof of Lemma 1}
	Let $\hat{\pmb{d}}=(q_1(\bar{X}_1-\bar{Y}_1),...,q_p(\bar{X}_p-\bar{Y}_p)) $. We have
	\begin{align*}
		\mathbb{E}\|\hat{\pmb{d}}-\pmb{d}\|^2&=\sum_{k=1}^{p}\left\{\mathbb{E}(q_k^2(\bar{X}_k-\bar{Y}_k)^2)-2d_k\mathbb{E}\{q_k(\bar{X}_k-\bar{Y}_k)\}+d_k^2    \right\}\\
		&=\sum_{k=1}^{p}\left\{ \text{Cov}\{q_k^2,(\bar{X}_k-\bar{Y}_k)^2\}+\mathbb{E}(q_k^2)\mathbb{E}(\bar{X}_k-\bar{Y}_k)^2
		-2d_k\left[ \text{Cov}\left\{q_k,(\bar{X}_k-\bar{Y}_k)\right\}+d_k\mathbb{E}q_k\right]+d_k^2  \right\}.
	\end{align*}
	Since $q_k$ is bounded between $0$ and $1$, $\text{Var}(q_k^2)$ and $\text{Var} (q_i)$ are both bounded. For $1\leq k \leq s_1$, note that $(\bar{X}_k-\bar{Y}_k)\sim N(d_k,\sigma_{kk}\nn) $, and $d^2_k<\infty$ we have $\text{Var}\{(\bar{X}_k-\bar{Y}_k)^2\}=O(\nn)$. Use Cauchy-Schwarz inequality, for $1\leq k\leq s_1$, each summand has absolute value bounded by
	\begin{align*}
		&O(\nn)+ \mathbb{E}\left\{q_k^2\left(\nn\sigma_{kk}+d_k^2   \right)\right\}-2d_k^2\mathbb{E}(q_k)+d_k^2-2d_k\text{Cov}(q_k,\bar{X}_k-\bar{Y}_k)\\
		=&O(\nn)+d_k^2\mathbb{E}(q_k^2)+d_k^2-2d_k^2\mathbb{E}(q_k)+O(d_k\nn)+O(\nn)\\
		=&O(\nn)+O(d_k\nn)+O\left\{d^2_k(1-\mathbb{E}(q_k))^2\right\}+O\left\{d_k^2(E(q^2_k)-(Eq_k)^2)\right\}\\
		=&O(\nn)+O(d_k\nn)+o(d^2_k).
	\end{align*}
	For $s_1+1\leq i\leq p$,
	$$
	\mathbb{E}\{q_k(\bar{X}_k-\bar{Y}_k)-d_k\}^2\leq d_k^2.
	$$

	Hence, by (A4)
	\begin{equation}\label{pfl}
		\mathbb{E}\|\hat{\pmb{d}}-\pmb{d}\|^2=O(s_1\nn)+o(\sum_{k=1}^{s_1}d^2_k)+O(\sum_{k=1}^{s_1}d_k\nn)+o(1).
	\end{equation}
	Since  $s_1\nn=O(\sum_{k=1}^{s_1}d^2_k)$ and by assumption of case 1, $\sum_{k=1}^{s_1}d^2_k=O(1)$ and $\sum_{k=1}^{s_1}d_k\nn=o(1)$ we have $\mathbb{E}\|\hat{\pmb{d}}-\pmb{d}\|^2=o(1)$. Now, 
	since  $\bar{\pmb X}+\bar{\pmb Y}$ and $\bar{\pmb X}-\bar{\pmb Y}$ are independent, and using Basu's Theorem we know that $\bar{\pmb X}+\bar{\pmb Y}$ and $\hat{\sigma}_{kk}$ are independent, thus $\bar{\pmb X}+\bar{\pmb Y}$ and $\hat{d}$ are independent and we have 
	\begin{align*}
		&\mathbb{E}(S^\pi_j-\hat{S}_j)^2\\
		=&\mathbb{E}\left\{\left(\pmb{W}_j-\dfrac{\pmb{\mu}_1+\pmb{\mu}_2}{2}\right)^\top\Sigma^{-1}\pmb{d}-\left(\pmb{W}_j-\dfrac{\bar{\pmb X}+\bar{\pmb Y}}{2}\right)^\top\hat{\Sigma}^{-1}\hat{\pmb{d}}\right\}^2\\
		=&\mathbb{E}\left\{\left(\pmb{W}_j-\dfrac{\pmb{\mu}_1+\pmb{\mu}_2}{2}\right)^\top\Sigma^{-1}(\pmb{d}-\hat{\pmb{d}})\right.+\left(\dfrac{\bar{\pmb X}+\bar{\pmb Y}}{2}-\dfrac{\pmb{\mu}_1+\pmb{\mu}_2}{2}\right)^\top\Sigma^{-1}\hat{\pmb{d}}+\left.\left(\pmb{W}_j-\dfrac{\bar{\pmb X}+\bar{\pmb Y}}{2}\right)^\top(\Sigma^{-1}-\hat{\Sigma}^{-1})\hat{\pmb{d}}\right\}^2\\
		=&O\left[\mathbb{E}\left\{\left(\pmb{W}_j-\dfrac{\pmb{\mu}_1+\pmb{\mu}_2}{2}\right)^\top\Sigma^{-1}(\pmb{d}-\hat{\pmb{d}})\right\}^2\right.
		+\mathbb{E}\left\{\left(\dfrac{\bar{\pmb X}+\bar{\pmb Y}}{2}-\dfrac{\pmb{\mu}_1+\pmb{\mu}_2}{2}\right)^\top\Sigma^{-1}\hat{\pmb{d}}\right\}^2\\
		+&\left. \mathbb{E}\left\{\left(\pmb{W}_j-\dfrac{\bar{\pmb X}+\bar{\pmb Y}}{2}\right)^\top(\Sigma-\hat{\Sigma}^{-1})\hat{\pmb{d}}\right\}^2\right]\\
		=&\uppercase\expandafter{\romannumeral1}+\uppercase\expandafter{\romannumeral2}+\uppercase\expandafter{\romannumeral3}.
	\end{align*}
	For term $\uppercase\expandafter{\romannumeral1}$, as in case 1 we have $||\pmb{d}||^2$ bounded:
	\begin{align*}
		\mathbb{E}\left\{\left(\pmb{W}_j-\frac{\pmb{\mu}_1+\pmb{\mu}_2}{2}\right)^\top\Sigma^{-1}(\pmb{d}-\hat{\pmb{d}})\right\}^2=O\left\{\mathbb{E}\left(\lambda_{max}(\Sigma^{-1})^2||\pmb{d}-\hat{\pmb{d}}||^2\right)\right\},
	\end{align*}  
	where $\lambda_{max}(\Sigma^{-1})$ is the largest eigenvalue of $\Sigma^{-1}$, which in our case is bounded by some constant.\\ 
	For term $\uppercase\expandafter{\romannumeral2}$, we have
	\begin{align*}
		&\mathbb{E}\left\{\left(\dfrac{\bar{\pmb X}+\bar{\pmb Y}}{2}-\frac{\pmb{\mu}_1+\pmb{\mu}_2}{2}\right)^\top\Sigma^{-1}\hat{\pmb{d}}\right\}^2\\
		=& O\left\{\mathbb{E}\left(\nn\lambda_{max}(\Sigma^{-1})||\pmb{d}-\hat{\pmb{d}}||^2\right)+\mathbb{E}\left(\nn\lambda_{max}(\Sigma^{-1})||\pmb{d}||^2\right)\right\}\\
		=&O\left\{\nn\mathbb{E}\left(\lambda_{max}(\Sigma^{-1})(||\pmb{d}-\hat{\pmb{d}}||^2+||\pmb{d}||^2)\right)\right\}.
	\end{align*}  
	For term $\uppercase\expandafter{\romannumeral3}$, we have
	\begin{align*}
		&\mathbb{E}\left\{\left(\pmb{W}_j-\frac{\bar{\pmb X}+\bar{\pmb Y}}{2}\right)^\top(\Sigma^{-1}-\hat{\Sigma}^{-1})\hat{\pmb{d}}\right\}^2\\
		=&\mathbb{E}\left[\hat{\pmb{d}}^\top(\Sigma-\hat{\Sigma}^{-1})^\top\left\{\left(1+\frac{1}{4}\nn\right)\Sigma^{-1}+\frac{1}{4}\pmb{d}\pmb{d}^\top\right\}(\Sigma^{-1}-\hat{\Sigma}^{-1})\hat{\pmb{d}}\right]\\
		\le& \mathbb{E}\bigg[\left\{\left(1+\frac{1}{4}\nn\right)\frac{1}{\epsilon_0}+\frac{1}{4}\max\{d_id_j,1\le i,j\le p\}\right\} 
		\hat{\pmb{d}}^\top(\Sigma^{-1}-\hat{\Sigma}^{-1})^\top(\Sigma^{-1}-\hat{\Sigma}^{-1})\hat{\pmb{d}}\bigg]\\
		=& O\bigg[\mathbb{E}\left\{||(\Sigma^{-1}-\hat{\Sigma}^{-1})||_2^2||\hat{\pmb{d}}||^2\right\}\bigg]\\
		=& O\bigg[\mathbb{E}\left\{||(\Sigma^{-1}-\hat{\Sigma}^{-1})||_2^2||\pmb{d}||^2\right\}\bigg]\\
		=&o(1).	
	\end{align*}   
	The last equality follows from assumption (A2) and the fact that $||\pmb{d}||^2<\infty$ in case 1. Hence,
	\begin{align*}
		\mathbb{E}\left( S^\pi_j- \hat{S}_j  \right)^2&= O\left\{\mathbb{E}\left(\|\hat{\pmb{d}}-\pmb{d}\|^2\right)\right\}+O(\nn)+o(1)=o(1).
	\end{align*}
	Next we ask under what conditions do we have $\mathbb{E}\left|\hat{T}^j-\Tor^j\right|^2\rightarrow 0 $.
	Let $\delta>0$ be some constant, applying Chebyshev's inequality, we have $P\left( |\hat{S}_j- S^\pi_j|>\delta\right)= O\left\{\mathbb{E}\left(|S^\pi_j- \hat{S}_j | \right)^2\right\}\rightarrow 0$. When                               $ |\hat{S}_j- S^\pi_j|\le \delta$, apply Cauchy-Schwartz inequality, we have:
	\begin{align*}
		&\mathbb{E}_{\left|{S}^\pi_j-\hat{S} _j\right|<\delta}\left|\exp(\hat{S}_j)-\exp(S^\pi_j)\right|^2\\
		\le& \mathbb{E}e^{2S^\pi_j+2\delta}\mathbb{E}\left(S^\pi_j- \hat{S}_j  \right)^2\\
		=&e^{O(\sum_{k=1}^{s_1}d^2_k)}\cdot o(1).
	\end{align*}
	Under the assumption of case 1 the above goes to 0. 
	Therefore, as $\hat{T}^j,{T}^j$ are bounded above by 1, we have:
	\begin{align*}\label{eq:T}
		\mathbb{E}(\hat{T}^j-{T}^j)^2&\le \mathbb{E}\left|\hat{T}^j-{T}^j\right|\\
		&\le 2P\left(\left| \hat{S}_j- S^\pi_j\right|>\delta\right)+ E_{\left| \hat{S}_j- S^\pi_j\right|<\delta}\left|\exp(\hat{S}_j)-\exp(S^\pi_j)\right|.
	\end{align*}
	The lemma follows.\qed

	\subsection{Proof of Lemma 2}
	Using the definitions of $\hat{V}_j$ and $V_j$, we can show that
	\begin{align*}
		\dfrac{1}{2}\left(\hat{V}_j - V_j\right)^2 
		&\leq \left(\hat{T}^{ j} - \Tor^{ j}\right)^2\mathbb{I} \left(\hat{T}^{ j} \leq t, \Tor^{ j} \leq t \right) + \left(\hat{T}^{ j}-\alpha_2\right)^2\mathbb{I} \left(\hat{T}^{ j} \leq t, \Tor^{ j}> t \right)\\
		&+ \left(\Tor^{ j}-\alpha_2\right)^2\mathbb{I} \left(\hat{T}^{ j} > t, \Tor^{ j} \leq t \right).
	\end{align*}
	Let us refer to the three summands on the right hand as $(1)$, $(2)$ and $(3)$ respectively. By Lemma 1, $(1) = o(1)$. Then let $\varepsilon > 0$, and consider that
	\begin{align*}
		&P\left(\hat{T}^{ j} \leq t, \Tor^{ j}> t \right) \\
		\leq& P\left(\hat{T}^{ j} \leq t, \Tor^{ j}\in \left(t, t+ \varepsilon \right) \right)+ P\left(\hat{T}^{ j} \leq t, \Tor^{ j}\geq  t+ \varepsilon  \right) \\
		\leq& P\left\{\Tor^{ j} \in \left(t, t+ \varepsilon \right)\right\} + P\left(\left|\Tor^{ j} -\hat{T}^{ j}\right| > \varepsilon \right).
	\end{align*}
	The first term on the right hand is vanishingly small as $\varepsilon \rightarrow 0$ because $\Tor^{ j}$ is a continuous random variable. The second term converges to $0$ by Lemma 1. Noting that $0\leq\Tor^{ j} \leq 1$, we conclude $(2) = o(1)$. In a similar fashion, we can show that $(3) = o(1)$, thus proving the lemma.\qed

	\section{Asymptotic Equivalence of FSR and mFSR}\label{fsrmsfr}
	We show $\text{FSR}_{\pmb{\delta}}^c$ and $\text{mFSR}_{\pmb{\delta}}^c$ are asymptotically equivalent. Let $\mathcal{X}_{\pmb{\delta}}^c=\frac{1}{m}\sum_{j=1}^{m}\mathbb{I}(\delta_j=c,\theta_j\neq c)$ and $\mathcal{Y}_{\pmb{\delta}}^c=\frac{1}{m}\sum_{j=1}^{m}\mathbb{I}(\delta_j=c)$. The goal is to show
	$
	|\text{FSR}_{\pmb{\delta}}^c-\text{mFSR}_{\pmb{\delta}}^c|=o(1).
	$
	\begin{align*}
		|\text{FSR}_{\pmb{\delta}}^c-\text{mFSR}_{\pmb{\delta}}^c|\leq \mathbb{E}\left\{\left|      \dfrac{\mathcal{X}_{\pmb{\delta}}^c}{\mathcal{Y}_{\pmb{\delta}}^c}-\dfrac{\mathcal{X}_{\pmb{\delta}}^c}{\mathbb{E}\mathcal{Y}_{\pmb{\delta}}^c}\right|\mathbb{I}(\mathcal{Y}_{\pmb{\delta}}^c>0)\right\}=\mathbb{E}\left\{\dfrac{\mathcal{X}_{\pmb{\delta}}^c}{\mathcal{Y}_{\pmb{\delta}}^c}\mathbb{I}(\mathcal{Y}_{\pmb{\delta}}^c>0)\dfrac{|\mathcal{Y}_{\pmb{\delta}}^c-\mathbb{E}\mathcal{Y}_{\pmb{\delta}}^c|}{\mathbb{E}\mathcal{Y}_{\pmb{\delta}}^c}\right\}
	\end{align*}
	Since $\mathcal{X}_{\pmb{\delta}}^c\leq \mathcal{Y}_{\pmb{\delta}}^c$, we have
	$$
	\mathbb{E}\left\{\dfrac{\mathcal{X}_{\pmb{\delta}}^c}{\mathcal{Y}_{\pmb{\delta}}^c}\mathbb{I}(\mathcal{Y}_{\pmb{\delta}}^c>0)\dfrac{|\mathcal{Y}_{\pmb{\delta}}^c-\mathbb{E}\mathcal{Y}_{\pmb{\delta}}^c|}{\mathbb{E}\mathcal{Y}_{\pmb{\delta}}^c}\right\}\leq \mathbb{E}\left\{   \dfrac{|\mathcal{Y}_{\pmb{\delta}}^c-\mathbb{E}\mathcal{Y}_{\pmb{\delta}}^c|}{\mathbb{E}\mathcal{Y}_{\pmb{\delta}}^c}  \right\}\leq \dfrac{(\mathbb{E}|\mathcal{Y}_{\pmb{\delta}}^c-\mathbb{E}\mathcal{Y}_{\pmb{\delta}}^c|^2)^{1/2}}{\mathbb{E}\mathcal{Y}_{\pmb{\delta}}^c}=\dfrac{  (\Var \mathcal{Y}_{\pmb{\delta}}^c)^{1/2}  }{\mathbb{E} \mathcal{Y}_{\pmb{\delta}}^c}
	$$
	By assumption (A3), we have $m\mathcal{Y}_{\pmb{\delta}_{OR}}^c\sim \text{Binom}(m,\eta)$ with $\eta>0$. 
	Therefore, $\mathbb{E}\mathcal{Y}^c_{\pmb{\delta}_{OR}}=\eta$ and
	$\Var(\mathcal{Y}^c_{\pmb{\delta}_OR})^{1/2}=\sqrt{\eta(1-\eta)/m}$,
	$	|\text{FSR}_{\pmb{\delta}_{OR}}^c-\text{mFSR}_{\pmb{\delta}_{OR}}^c| =O(m^{-1/2})$. In the setting of Theorem  \ref{thm3}, the data driven procedure is mimicking $\pmb{\delta}_{OR}$. It can be seen from the proof of Theorem \ref{thm3} that $\text{mFSR}$ and $\text{FSR}$ of the data driven procedure are also asymptotically equivalent. Similarly, for the setting of Theorem \ref{thm4}, consider the rule $\pmb{\delta}_Z$ defined in the proof of Theorem \ref{thm4}, by assumption (A3) we also have $	|\text{FSR}_{\pmb{\delta}_{Z}}^c-\text{mFSR}_{\pmb{\delta}_{Z}}^c| =O(m^{-1/2})$. In the setting of Theorem  \ref{thm4}, the data driven procedure is mimicking $\pmb{\delta}_{Z}$. It can be seen from the proof of Theorem \ref{thm4} that $\text{mFSR}$ and $\text{FSR}$ of the data driven procedure are also asymptotically equivalent.\qed

\end{document}